\documentclass[smallextended]{svjour3} 
\smartqed

\usepackage{mathrsfs}
\usepackage{graphicx}

\newcommand{\diff}{\mathrm{d}}

\begin{document}

\bibliographystyle{plain}

\title{Profiles of emission lines generated by rings orbiting braneworld Kerr black holes}
\author{Jan Schee and Zden\v{e}k Stuchl\'{\i}k}

\institute{J. Schee \at Institute of Physics, Faculty of Philosophy and Science, Silesian University in Opava, Bezru\v{c}ovo n\'{a}m. 13, 
CZ-746 01 Opava, Czech Republic\\
\email{schee@email.cz}
\and
Z. Stuchl\'{i}k \at
Institute of Physics, Faculty of Philosophy and Science, Silesian University in Opava, Bezru\v{c}ovo n\'{a}m. 13, 
CZ-746 01 Opava, Czech Republic\\
\email{Zdenek.Stuchlik@fpf.slu.cz}
}

\maketitle

\begin{abstract}
In the framework of the braneworld models, rotating black holes can be described by the Kerr metric with a tidal charge representing the influence of the non-local gravitational (tidal) effects of the bulk space Weyl tensor onto the black hole spacetime. We study the influence of the tidal charge onto profiled spectral lines generated by radiating tori orbiting in vicinity of a rotating black hole. We show that with lowering the negative tidal charge of the black hole, the profiled line becomes to be flatter and wider keeping their standard character with flux stronger at the blue edge of the profiled line. The extension of the line grows with radius falling and inclination angle growing. With growing inclination angle a small hump appears in the profiled lines due to the strong lensing effect of photons coming from regions behind the black hole. For positive tidal charge ($b>0$) and high inclination angles two small humps appear in the profiled lines close to the red and blue edge of the lines due to the strong lensing effect. We can conclude that for all values of $b$, the strongest effect on the profiled lines shape (extension) is caused by the changes of the inclination angle.

\keywords{black hole, emission line, braneworld model}
\PACS{97.60.Lf, 95.30.Sf}
\end{abstract}

\section{Introduction}
String theory and M-theory describing gravity as a truly
higher-dimensional interaction becoming effectively 4D at low-enough energies inspired studies of the braneworld models, where the
observable universe is a 3-brane (domain wall) to which the
standard model (non-gravitational) matter fields are confined,
while gravity field enters the extra spatial dimensions the size
of which may be much larger than the Planck length scale
$l_\mathrm{P}\sim 10^{-33}\, \mathrm{cm}$ \cite{Ark-Dim-Dva:1998:}.
Gravity can be localized near the brane at low energies even with a non-compact, infinite size extra dimension with the warped spacetime satisfying the 5D Einstein equations with negative cosmological constant \cite{Ran-Sun:1999:}.

Significant deviations from the Einstein gravity occur at very high energies in vicinity of compact objects (see e.g., \cite{Maa:2004:,Ger-Maa:2001:,Ali-Gum:2005:}). The high-energy effects produced by the gravitational collapse are disconnected from the outside space by the horizon, but they could have a signature on the brane, influencing properties of black holes \cite{Maa:2004:}. There are high-energy effects of local character influencing pressure in collapsing matter  The non-local corrections of "back-reaction`` character arise from the influence of the Weyl curvature of the bulk space on the brane - the matter on the brane induces  Weyl curvature in the bulk which makes influence on the structures on the brane due to the bulk gravitation stresses \cite{Maa:2004:,Dad-etal:2000:}. The combination of high-energy (local) and bulk stress (non-local) effects alters significantly the matching problem on the brane, as compared to the 4D Einstein gravity; for spherical objects, matching no longer leads to a Schwarzchild exterior in general \cite{Dad-etal:2000:,Ger-Maa:2001:}.  The Weyl stresses induced by bulk gravitons imply that the matching conditions do not have unique solution on the brane; in fact, knowledge of the 5D Weyl tensor is needed as a minimum condition for uniqueness \cite{Ger-Maa:2001:}. However, recently no exact 5D solution is known. Assuming spherically symmetric metric induced on the 3-brane the effective gravitational field equations of vacuum type in both the brane and bulk can be solved, giving a Reissner-Nordstr\"{o}m static black hole solutions endowed with a "tidal'' charge parameter $b$ \cite{Dad-etal:2000:}, instead of the standard electric charge parameter $Q^2$ \cite{MTW}. The tidal charge reflects the effects of the Weyl curvature of the bulk space, i.e., from the 5D gravitation stresses \cite{Dad-etal:2000:,Maa:2004:} with bulk gravitation tidal effects giving the name of the charge. Note that the tidal charge can be both positive and negative, and there are some indications that the negative tidal charge should properly represent the ``back-reaction`` effects of the bulk space Weyl tensor on the \cite{Maa:2004:,Dad-etal:2000:,Sasaki:2000:} 

The stationary and axisymmetric solution of the constrained equations describing
rotating black holes localized in the Randall-Sundrum braneworld
were derived in \cite{Ali-Gum:2005:}. The solutions are determined by metric tensor of the
Kerr-Newman form with a tidal charge describing the 5D correction
term generated by the 5D Weyl tensor stresses. The tidal charge
has an ``electric'' character and arises due to the 5D
gravitational coupling between the brane and the bulk, reflected
on the brane through the ``electric'' part of the bulk Weyl tensor
\cite{Ali-Gum:2005:}, in analogy with the spherically
symmetric black-hole  case \cite{Dad-etal:2000:}.
When the electromagnetic field is introduced, the non-vacuum solutions of the effective Einstein equations on the brane are much more complex in comparison with the standard Kerr-Newman solutions \cite{Ali-Gum:2005:}.

Here we consider optical phenomena  in the Kerr-Newman type of solutions describing the braneworld rotating (Kerr) black holes with no
electric charge, since in astrophysically relevant situations the
electric charge of the black hole must be exactly zero, or very
small \cite{MTW}. Then the results obtained in analyzing the
behavior of test particles and photons or test fields around the
Kerr-Newman black holes could be used assuming both positive and
negative values of the braneworld tidal parameter $b$ (used instead of
the charge parameter $Q^2$).

The information on the properties of strong gravitational fields in vicinity of compact objects, namely
of black holes, is encoded into optical phenomena of different kind that enable us to make estimates of the black hole parameters, including its tidal charge, when predictions of the theoretical models are confronted with the observed data. From this point of view, the spectral profiles of accretion discs around the black holes in galactic binaries, e.g., in microquasars, are most promising \cite{McCli-Nar-Sha:2007:}, along with profiled spectral lines in the X-ray flux \cite{Bao-Stu:1992:,Stu-Bao:1992:,Laor:1991:,Mat-Fab-Ros:1993:,Viergutz:1993:,Rau-Bla:1994:,Fan-Cal-Fel-Cad:1997:,Zak:2003:,Zak-Rep:2006:,Czerny_etal:2007:}. Important information could also be obtained from the quasiperiodic oscillations observed in the X-ray flux of some low-mass black hole binaries of Galactic origin \cite{Rem:2005:,Rem-McCli:2006:ARASTRA:}, some expected intermediate black hole sources \cite{Strohmayer:2007:}, or those observed in Galactic nuclei \cite{Asch:2004:ASTRA:,Asch:2007:}. The most promising orbital resonance model then enables relative exact measurement of the black hole parameters \cite{Tor-Abr-Klu-Stu:2005:,Tor:2005a:,Tor:2005b:} that should be confronted with the predictions of the optical modelling \cite{McCli-Nar-Sha:2007:}. In the case of our Galaxy centre black hole Sgr A$^*$, we could be able to measure the detailed optical phenomena, as compared with the other sources, since it is the nearest supermassive black hole with mass estimated to be $\sim 4\times 10^6 M_\odot$ \cite{Ghe-etal:2005:}, enabling us to measure the "silhouette`` of the black hole and other subtle GR phenomena in both weak and strong field limits \cite{Cun-Bar:1973:,SS:a:RAGTime:2007:Proceedings,SSJ:RAGTime:2005:Proceedings}.

Here we present an introductory study on the role of the braneworld tidal charge parameter in the optical phenomena related to profiled spectral lines generated by thin radiating tori in the braneworld Kerr black-hole backgrounds. Of course, for $b>0$, the results hold as well for standard Kerr-Newman spacetimes due to the correspondence $b\rightarrow Q^2$ with $Q^2$ being the squared charge parameter of the Kerr-Newman black hole. We use here the transfer function method \cite{Bao-Stu:1992:} which seems to be the most appropriate in our problem. In section 2, the geometry and photon equations of motion are given. In section 3, the geodetic Al circular orbits in the equatorial plane of the Kerr spacetimes with a tidal charge are presented. In section 4, the specific energy flux of photons is given and optical phenomena related to the thin ring of radiating particles are discussed, namely the frequency shift and the focusing. In section 5, the profiled spectral lines are calculated using the transfer function method and assuming isotropically radiating sources with photon energy fixed to $E_0$ in the rest frame of the radiating source. The line profiles are determined in dependence on the viewing angle $\theta_0$ and the ring radius $r_e$, which is assumed to be located in the inner part of a thin accretion disc. Specially, it is assumed near marginally stable orbit $r_{ms}$ and in the orbit corresponding to the resonance radius $r_{3:2}$ where the ratio of vertical and radial epicyclic frequencies is $\sim 3/2$, corresponding to the QPOs frequency ratios commonly observed in microquasars \cite{Tor-Abr-Klu-Stu:2005:,Tor:2005b:}. In Section 6, concluding remarks are presented.

\section{The geometry of braneworld Kerr spacetime and equations of motion for test particles and photons}
The solution of vacuum effective Einstein equations on the brane has been for rotating black holes given in \cite{Ali-Gum:2005:}. It is not an exact metric satisfying the full system of the 5D equations (which is not known recently), but a consistent solution of constraint equations with a specialized metric form  on the brane. In such a framework it represents the only useful rotating black hole solution reflecting the influence of the extra-dimensions by a single parameter called braneworld tidal charge because of expressing the tidal effects of the bulk \cite{Ali-Gum:2005:}.  The properties of the circular motion in braneworld Kerr spacetimes were discussed in \cite{Ali-Gum:2005:,Stu-Kot:2007:}. We briefly summarize the relevant results.

\subsection{Effective Gravitational Equations on the brane}

The 5D Einstein equations in the bulk spacetime have the form \cite{Shi-Mae-Sas:2000:,Dad-etal:2000:}

\begin{equation}
 \phantom{I}^{(5)}G=\phantom{I}^{(5)}R_{AB}-\frac{1}{2}g_{AB}\phantom{I}^{(5)}R=-\Lambda_5 g_{AB}+\kappa_5^2\left(\phantom{I}^{(5)}T_{AB}+\sqrt{\frac{h}{g}}\tau_{AB}\delta Z\right),
\end{equation}
where $\kappa_5^2=8\pi G_5$ ($G_5$ being the gravitational constant), $\Lambda_5$ is the bulk cosmological constant (assuming anti-de Sitter geometry), $\phantom{I}^{(5)}T_{AB}$ is the energy-momentum tensor in the bulk, $\tau_{AB}$is the energy-momentum tensor on the brane, $h$ and $g$ being metric determinants of $h_{\alpha\beta}$ and $g_{AB}$, and the bulk spacetime is expressed in the form
\begin{equation}
 g_{AB}=n_A n_B + h_{\alpha\beta}e^\alpha_A e^\beta_B.
\end{equation}
Here, $n_A$ is the unit vector orthogonal to the brane and $e_A^\alpha$ represents local frame of four-vectors;  the induced metric on the brane 

\begin{equation}
 h_{\alpha\beta} = g_{AB}e_\alpha^A e_\beta^B.
\end{equation}

The effective Einstein gravitational equations (EGE) on the brane could then be given by using the Israel junction method generalized to 5D situation. The gravitational field equations on the brane take the form \cite{Shi-Mae-Sas:2000:}

\begin{equation}
 G_{\alpha\beta}=-\Lambda h_{\alpha\beta} \kappa_4^2 T_{\alpha\beta}+ \kappa_5^4 S_{\alpha\beta} -W_{\alpha\beta}-3\kappa_5^2 U_{\alpha\beta}.\label{EGE}
\end{equation}
The traceless tensor

\begin{equation}
	W_{\alpha\beta}=A_{\alpha\beta}-\frac{1}{4}h_{\alpha\beta}A 
\end{equation}
is constructed from the "electric'' part of the bulk Riemann tensor

\begin{equation}
 	A_{\alpha\beta}=\phantom{I}^{(5)}R_{ABCD}n^An^C e_\alpha^B e_\beta^D,\quad A=h_{\alpha\beta}A^{\alpha\beta}.
\end{equation}
The cosmological constant on the brane

\begin{equation}
 \Lambda = \frac{1}{2}\left(\Lambda_5+\frac{1}{6}\kappa_5^4\lambda^2-\kappa_5^2 P\right),\label{LAMBDA}
\end{equation}
where

\begin{equation}
 P=\phantom{I}^{(5)}T_{AB}n^A n^B
\end{equation}
is the normal compressive pressure term in the 5D spacetime, and 4D gravitational constant is related to the brane tension $\lambda$ by the relation

\begin{equation}
 \kappa_4^2=\frac{1}{6}\kappa_5^4\lambda.\label{KAPPA}
\end{equation}
The  "squared energy-momentum`` tensor is given by

\begin{equation}
 S_{\alpha\beta}=-\frac{1}{4}\left[\left(T_\alpha^\gamma T_{\gamma\beta}-\frac{1}{3}TT_{\alpha\beta}\right)-\frac{1}{2}h_{\alpha\beta}\left(T_{\gamma\delta}T^{\gamma\delta}-\frac{1}{3}T^2\right)\right],
\end{equation}
while the traceless braneworld part of the bulk energy-momentum tensor is 

\begin{equation}
 U_{\alpha\beta}=-\frac{1}{3}\left(\phantom{I}^{(5)}T_{\alpha\beta}-\frac{1}{4}h_{\alpha\beta}h^{\gamma\delta}\phantom{I}^{(5)}T_{\gamma\delta}\right).\label{BulkEnMomTen}
\end{equation}

In the effective 4D EGE (\ref{EGE})-(\ref{BulkEnMomTen}), $W_{\alpha\beta}$ describes non-local gravitational effects of the bulk space onto the brane and is sometimes called Weyl fluid, while the local bulk effects on the brane are given by $S_{\alpha\beta}$, $U_{\alpha\beta}$ and $P$.

It should be stressed that the self-consistent solutions of the effective 4D EGE (\ref{EGE})-(\ref{BulkEnMomTen}) on the brane require the knowledge of the non-local gravitational and energy-momentum terms coming from the bulk spacetime. Therefore, the braneworld field equations are not closed in general and evolution equations into the bulk have to be solved for the projected bulk curvature and energy-momentum  tensors. However, in particular cases the braneworld-equations  system could be made closed assuming a special ansatz for the induced metric. In this way, both spherically symmetric \cite{Dad-etal:2000:} and axially symmetric braneworld black hole spacetimes \cite{Ali-Gum:2005:} have been found. Assuming vacuum bulk and braneworld spacetimes, the non-local gravitational effects of the bulk could be simply represented by the so called tidal charge entering the standard metric of the black hole spacetimes.

The rotating black holes localized on a 3-brane in the Randall-Sundrum braneworld model were derived under the assumption of stationary and axisymmetric Kerr-Schild metric on the brane and supposing empty bulk space and no matter fields on the brane ($T_{\alpha\beta}=0$) \cite{Ali-Gum:2005:}. The specialized solution of the constrained equations is thus assumed in the form

\begin{equation}
 \diff s^2 = \eta_{\mu\nu}\diff x^\mu\diff x^\nu + H(l_i\diff x^i)^2.
\end{equation}
The effective 4D Einstein equations then reduce to the form

\begin{equation}
 R_{\alpha\beta}=-E_{\alpha\beta},
\end{equation}
where

\begin{equation}
 E_{\alpha\beta}=\phantom{I}^{(5)}C_{ABCD}n^A n^B e_\alpha^C e_\beta^D
\end{equation}
is the $electric$ part of the 5D Weyl tensor, used besides the $W_{\alpha\beta}$ tensor to describe the non-local gravitational effects of the bulk space onto the brane. Further, the relations

\begin{equation}
 \Lambda_5=-\frac{6}{l^2},\quad G_4=\frac{G_5}{l}
\end{equation}
can be deduced from Eqs (\ref{LAMBDA}) and (\ref{KAPPA}), assuming zero cosmological constant on the brane ($\Lambda_4=0$). Here,

\begin{equation}
 l=\frac{6}{\lambda\kappa_5^2}
\end{equation}
is the curvature radius of the anti-de Sitter spacetime. (Henceforth we set $G_4=1$.)

\subsection{Geometry}
Using the standard Boyer-Lindquist coordinates ($t$, $r$, $\theta$, $\varphi$) and geometric units (c=G=1) we can write the line element of Kerr black-hole metric on the 3D-brane  in the form \cite{Ali-Gum:2005:} 
\begin{eqnarray}
	\diff s^2 &=& -(1-\frac{2Mr - b}{\Sigma})\diff t^2 + \frac{\Sigma}{\Delta}\diff r^2 + \Sigma \diff \theta^2 + \frac{A}{\Sigma}\diff\varphi^2\nonumber\\
	&& - 2\frac{2Mr-b}{\Sigma}\sin^2\theta\diff t\diff\phi,\label{eq1}
\end{eqnarray}
where 

\begin{eqnarray}
	\Sigma &=& r^2 + a^2\cos^2\theta\label{eq2}\\
	\Delta &=& r^2 - 2Mr + a^2 +b\label{eq3}\\
	A &=& (r^2 + a^2)^2 - a^2\Delta\sin^2\theta\label{eq4}.
\end{eqnarray}
M is the mass parameter, $a=J/M$ is the specific angular momentum, the braneworld parameter $b$ is called \emph{tidal charge} and represents the imprint of non-local (tidal) gravitational effects from the bulk space \cite{Dad-etal:2000:}. The form of the metric (\ref{eq1}) is the 
same as in the case of the Kerr-Newman metric, with the tidal charge being replaced by the squared electric charge, $Q^2$ \cite{MTW}. 
The stress tensor on the brane $E_{\mu\nu}$ takes the form

\begin{eqnarray}
 E_t^{\phantom{t}t}&=&-E_\varphi^{\phantom{\varphi}\varphi}=-\frac{b}{\Sigma^3}[\Sigma-2(r^2+a^2)]\\
 E_r^{\phantom{r}r}&=&-E_\theta^{\phantom{\theta}\theta}=-\frac{b}{\Sigma^2}\\
 E_\varphi^{\phantom{\varphi}t}&=&-(r^2+a^2)\sin^2\theta E_t^{\phantom{t}\varphi} =-\frac{2ba}{\Sigma^3}(r^2+a^2)\sin^2\theta
\end{eqnarray}
that is fully analogical ($b\rightarrow Q^2$) to the components of electromagnetic energy-momentum tensor for the Kerr-Newman spacetimes in Einstein's general relativity \cite{Ali-Gum:2005:}. For simplicity, we put $M=1$ in the following, using thus dimensionless coordinates and parameters. The tidal charge can be both positive and negative; however, there are some indications favorizing negative values of $b$ \cite{Dad-etal:2000:,Maa:2004:}. Note that in the case of $b<0$, some braneworld black holes have the ring singularity of spacelike character \cite{Dad-etal:2000:}.

\subsection{Carter's equations}
In order to study the optical effects in braneworld Kerr spacetimes we have to solve  equations of motion of  photons. It is well known that photons move along null geodesics of the spacetime under consideration. 

Using the Hamilton-Jacobi method, Carter found separated first order differential equations of motion \cite{Carter:1968:}, which in the case of  the braneworld Kerr spacetimes read

\begin{eqnarray}
 	\Sigma\frac{\diff r}{\diff w}&=&\pm\sqrt{R(r)},\label{ce7}\\
 	\Sigma\frac{\diff \theta}{\diff w}&=&\pm\sqrt{W(\theta)},\label{ce8}\\
 	\Sigma\frac{\diff \varphi}{\diff w}&=&-\frac{P_W}{\sin^2\theta}+\frac{a P_R}{\Delta},\label{ce9}\\
 	\Sigma\frac{\diff t}{\diff w}&=&-a P_W + \frac{(r^2+a^2)P_R}{\Delta},\label{ce10}
\end{eqnarray}
where 

\begin{eqnarray}
  R(r)&=&P^2_R-\Delta(m^2 r^2 + K),\label{ce11}\\
  W(\theta)&=&(K-a^2m^2\cos^2\theta)-\left(\frac{P_w}{\sin\theta}\right)^2,\label{ce12}\\
  P_R(r)&=&E(r^2+a^2)-a\Phi,\label{ce13}\\
  P_W(\theta)&=&aE\sin^2\theta - \Phi.\label{ce14}
\end{eqnarray}
$E$ is the energy, $\Phi$ is the axial angular momentum and $K$ is the constant of motion related to total angular momentum that is usually replaced by the constant $Q=K-(aE-\Phi)^2$, since for motion in the equatorial plane ($\theta=\pi/2$) there is $Q=0$. For motion of photon, we put $m=0$. Generally, these equations can be integrated and expressed in terms of elliptic integrals \cite{SS:a:RAGTime:2007:Proceedings}.  
The analysis of photon motion in the Kerr-Newman spacetimes \cite{Stu:1981b:} can be directly applied to the case of photon motion in the braneworld Kerr spacetimes. It is done  in \cite{SS:a:RAGTime:2007:Proceedings}; we shall use the results in the following. As a source of radiation we consider here radiating rings composed of particles following equatorial circular geodesics. It is widely assumed that the observed profiled spectral lines of X-rays corresponding usually to the $Fe$ line at the rest energy $E_0\sim 6.7$keV are produced in the internal parts of accretion disc near the marginally stable circular geodesic representing roughly the inner edge of the disc \cite{Bao-Stu:1992:,Mat-Fab-Ros:1993:}. Therefore we briefly summarize properties of circular geodesics and the marginally stable orbits in dependence on the tidal charge $b$.

\section{Rotating ring in the equatorial plane of braneworld Kerr black hole}
Following the Carter equations for the circular motion at a given radius $r$ in the equatorial plane ($\theta=\pi/2$, $\diff\theta/\diff w = 0$), we find, solving simultaneously relations $R(r)=0$, $\diff R/\diff r = 0$,  the specific energy and angular momentum related to mass $m$ of the orbiting particle to be given by \cite{Ali-Gum:2005:,Stu-Kot:2007:}
\begin{eqnarray}
 E_\pm &=&\frac{r^2-2r+b\pm a\sqrt{r-b}}{Z_\pm},\label{energy}\\
 \Phi_\pm &=&\pm\frac{(r^2+a^2)\sqrt{r-b}\mp a(2r-b)}{Z_\pm},\label{ang_momentum}
\end{eqnarray}
where

\begin{equation}
 Z_\pm=r\sqrt{r^2-3r+2b\pm 2a\sqrt{r-b}};
\end{equation}
the upper (lower) sign corresponds to the corotating (counterrotating) orbits. The  crucial limiting radii for existence of circular orbits correspond to the photon circular orbits given by the real positive roots of the equations

\begin{equation}
 Z_\pm = 0,\quad r^2-3r +2b\pm 2a\sqrt{r-b}=0. 
\end{equation}
The marginally stable orbits ($\diff^2 R/\diff r^2 = 0$) are given by the relation \cite{Ali-Gum:2005:}

\begin{equation}
	r(6r - r^2 - 9b + 3a^2)+4b(b-a^2)\mp 8a(r-b)^{3/2}=0.
\end{equation}

\begin{figure}
 \begin{center}
\begin{tabular}{cc}
  \includegraphics[width=5cm]{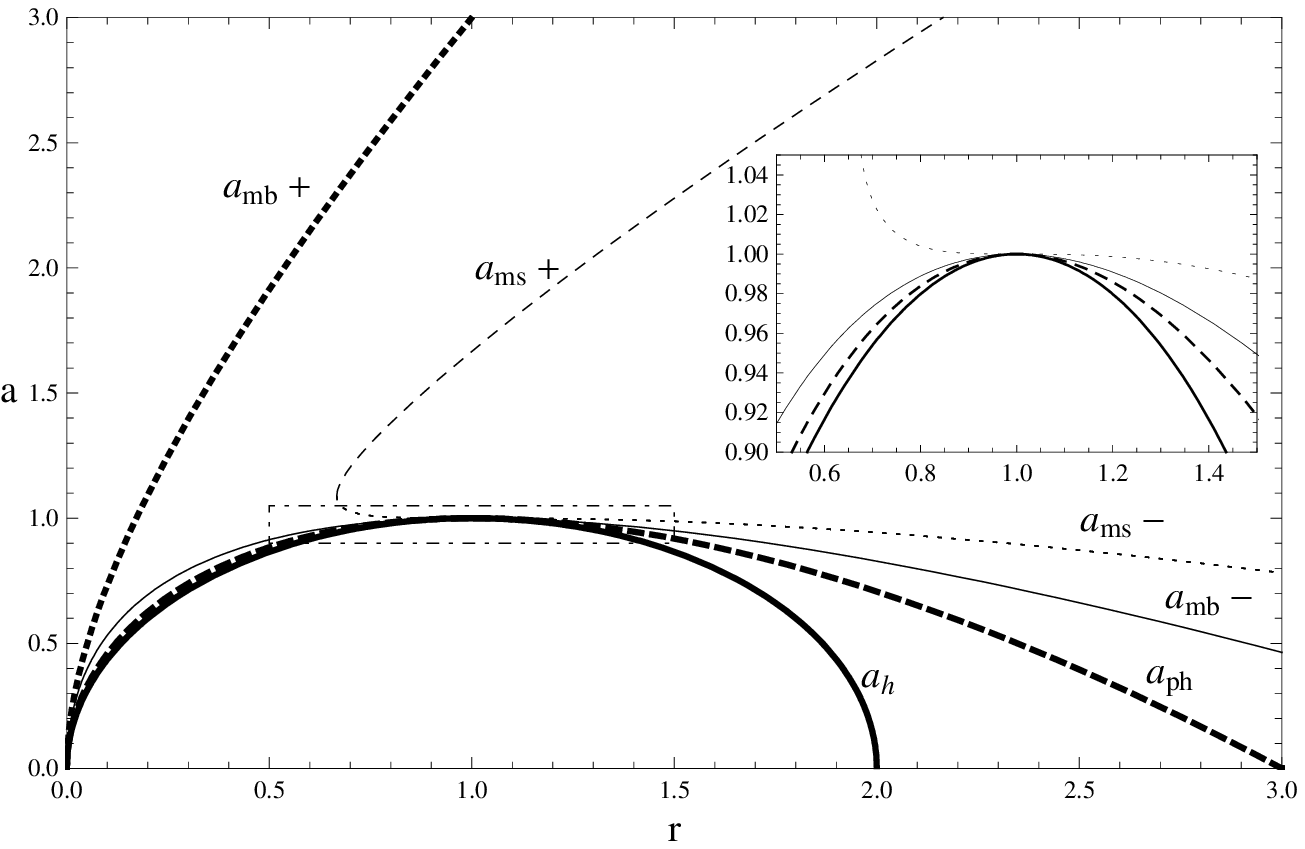}&\includegraphics[width=5cm]{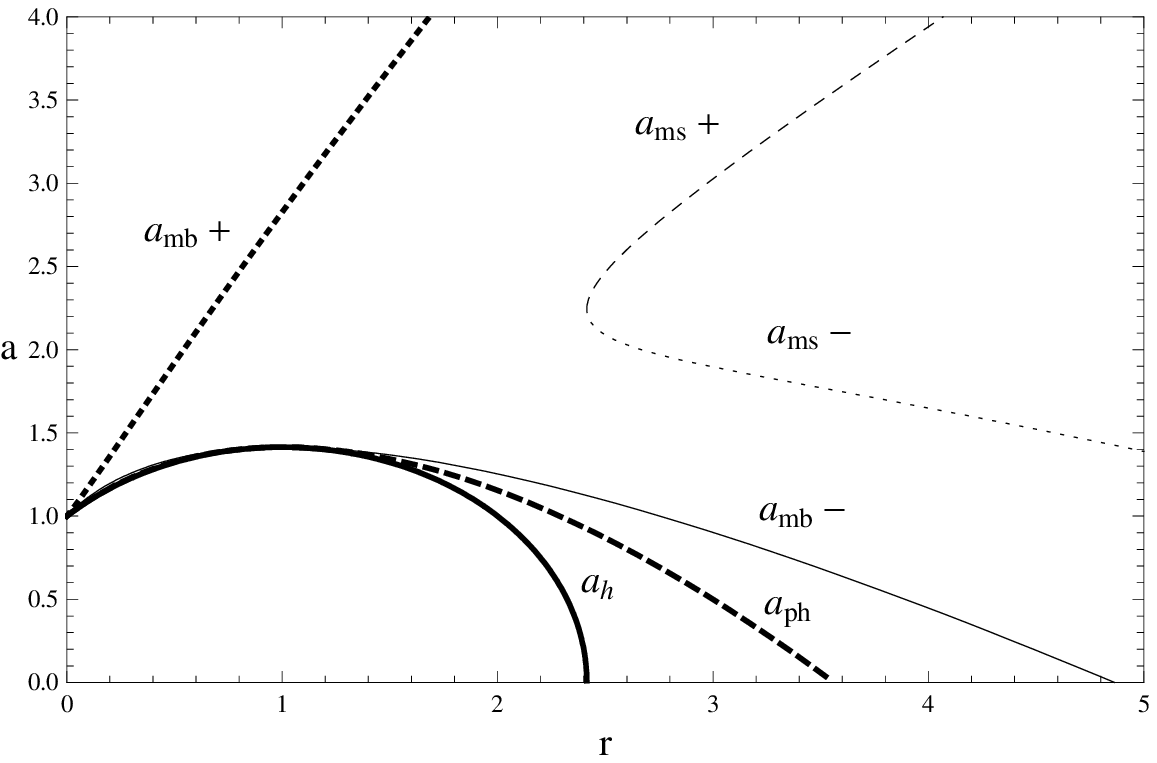}
\end{tabular}
\begin{tabular}{c}
  \includegraphics[width=5cm]{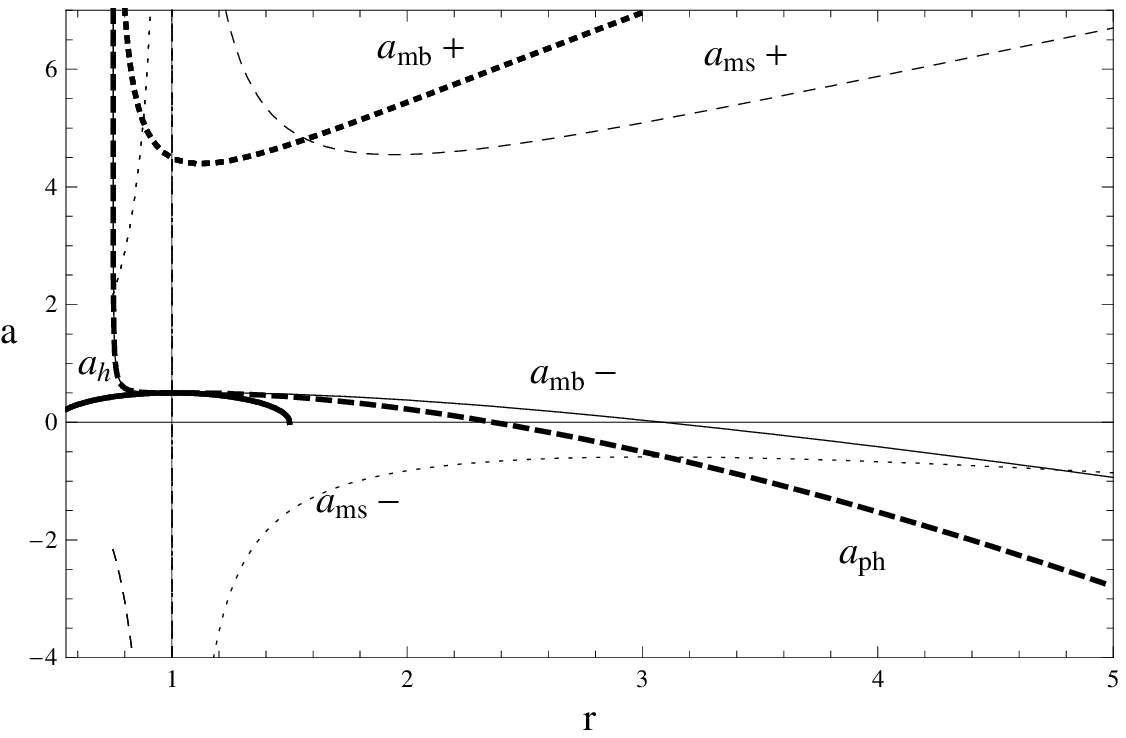}
\end{tabular}
 \end{center}
\caption{The plots of functions $a_h$, $a_{ph}$, $a_{ms\pm}$, $a_{mb\pm}$ for three representative values of tidal charge $b=0$ (top), $-1$ (middle), $+0.75$ (bottom).}\label{obr1}
\end{figure}

For fixed dimensionless tidal charge $b$, it is useful to express the loci of the horizons, photon circular geodesics, marginally bound orbits (with $E=1$) and marginally stable orbits by the following relations

\begin{eqnarray}
 a&=&a_h(r,b)\equiv\sqrt{2r-r^2+b},\nonumber\\
a&=&a_{ph}(r,b)\equiv\frac{r(3-r)-2b}{2\sqrt{r-b}}\nonumber\\
a&=&a_{mb}(r,b)\equiv\frac{\sqrt{r-b}(2r-b\mp r\sqrt{r})}{r-b}\nonumber\\
a&=&a_{ms}(r,b)\equiv\frac{4(r-b)^{3/2}\mp r\sqrt{3r^2-2r(1+2b)+3b}}{3r-4b}
\end{eqnarray}
Their behavior is illustrated for typical values of $b$ in Fig \ref{obr1}. Clearly, the condition $r>b$ must be considered that could be relevant under the inner horizon or in naked singularity spacetimes.

We assume a bright rotating ring of particles following circular orbit assumed to be located above the photon circular orbit or marginally stable orbit, respectively. The ring is composed of a large number of monochromatically radiating point sources which move along the circular orbit at the radial distance $r_e$ and which radiate isotropically in their rest frame. The angular velocity $\Omega = \diff\varphi/\diff t$ of such sources as measured by distant observers is given by \cite{Stu-Kot:2007:}

\begin{equation}
 \Omega = \frac{\diff\varphi}{\diff t} = \frac{\sqrt{r-b}}{r^2 + a\sqrt{r-b}}. \label{omega}
\end{equation}

\section{Optical effects governing profiled spectral lines}
We consider simple situation when the radiation emitted by the orbiting ring has negligible interaction with intervening media (of the disc itself or some others) while traveling to the observer. Therefore, we shall use the standard method of transfer function \cite{Cun-Bar:1973:,Bao-Stu:1992:}, where the emission region and the observer area are related by the optical phenomena that we summarize below.
\subsection{The specific energy flux of photons}
The observed specific energy flux is given by the formula
\begin{figure}[!ht]
\begin{center}
 \includegraphics[width=10cm]{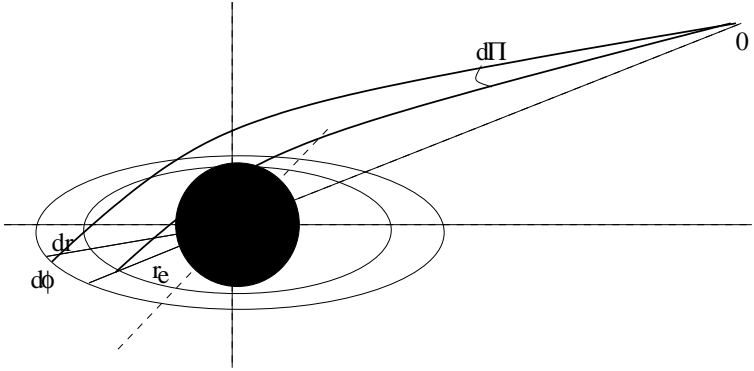}
\end{center}
\caption{The emitter $E$ emits isotropically in its rest frame. A photon is radiated at a directional angle $\delta$. It is received by observer $O$ at infinity. The coordinates of the received photon on the observer sky are [$\alpha$,$\beta$]. For details see \cite{SS:b:RAGTime:2007:Proceedings}}\label{obr2}
\end{figure}
\begin{equation}
 	F_o(\nu) = \int{I_o\diff\Pi},\label{flux_int}
\end{equation}
 where $I_o$ is observed specific intensity of the source and $\diff \Pi$ is the solid angle subtended by the source on the observer sky (see Fig.\ref{obr2}). The observed specific intensity and its value at the rest frame of source is given by the Liouville theorem

\begin{equation}
  \frac{I_o}{\nu_o^3}=\frac{I_e}{\nu_e^3}=const,
\end{equation}
where index $o$ ($e$) refers to the observer (emitter), $\nu_o$ ($\nu_e$) is the observed (emitted) frequency of photons. The observed specific flux then takes the form

\begin{equation}
 	F_o(\nu) = \int{I_e g^3 \diff\Pi},\label{flux}
\end{equation}
where we introduced the frequency shift ratio of the observed frequency to the emitted one, $g=\nu_o/\nu_e$. 

The frequency shift and the focusing of the bundle of rays are the key phenomena that form the profile of the observed spectral line.

\subsection{Frequency shift} 

The frequency shift $g$ can be expressed as the ratio of observed photon energy $E_o$ to emitted photon energy $E_e$
\begin{equation}
 g=\frac{E_o}{E_e}=\frac{(k_0)_\mu u_o^\mu}{(k_e)_\mu u_e^\mu},
\end{equation}
where $(k_0)_\mu$ ($(k_e)_\mu$) are covariant components of photon 4-momentum taken at the event of observation (emission) and $u^\mu_o$($u^\mu_e$) are contravariant components of the 4-velocity of the observer (emitter). In the case of static distant observer the 4-velocity reads $\mathbf{u}_o=(1,0,0,0)$. In the case of  emitter following a circular geodesic at $r=r_e$ in the equatorial plane of the braneworld Kerr black hole, the 4-velocity reads  $\mathbf{u}_e=(u_e^t,0,0,u_e^\varphi)$ with components being given by

\begin{eqnarray}
  u_e^t&=&\left[1-\frac{2}{r_e}(1-a\Omega)^2-(r_e^2+a^2)\Omega^2+\frac{b}{r_e^2}(1-2a\Omega)\right]^{-1/2},\\
u_e^\varphi&=&\Omega u_e^t,
\end{eqnarray}
where $\Omega$ is the angular velocity of the emitter as seen by distant observer and is given by Eq.(\ref{omega}). The total frequency shift, including both the gravitational and Doppler shifts, is given by

\begin{equation}
 g=\frac{\left[1-\frac{2}{r_e}(1-a\Omega)^2-(r_e^2+a^2)\Omega^2+\frac{b}{r_e^2}(1-2a\Omega)\right]^{1/2}}{1-\lambda\Omega},\label{freq_shift}
\end{equation}
where $\lambda=-k_\varphi/k_t$ is the impact parameter of the photon being a constant of the photon motion; notice that in our case of equatorial sources $g$ is explicitly independent of the second motion constant $q$. Of course, depending on the position of the emitter along the circular orbit, the motion constant of photons reaching a fixed distant observer will change periodically (see, e.g., \cite{Bao-Stu:1992:}).

\subsection{Focusing}
For a distant observer located at the distance $d_0$ from the source the solid angle $\diff\Pi$ can be expressed in terms of observer's plane coordinates $[\alpha,\beta]$ 

\begin{equation}
 \diff\Pi = \frac{1}{d^2_o}\diff\alpha\diff\beta.
\end{equation}
The observer's plane coordinates are in relation with photon's impact parameters $\lambda$ and $q$ given by formulae \cite{Cun-Bar:1973:,Bardeen:1973:}

\begin{equation}
 \alpha = -\frac{p^{(\phi)}}{p^{(t)}} = -\frac{\lambda}{\sin\theta_o},\label{alpha}
\end{equation}
and
\begin{equation}
 \beta =\frac{p^{(\theta)}}{p^{(t)}} =  \pm\sqrt{q^2+a^2\cos^2\theta_o-\lambda^2\cot^2\theta_o};\label{beta}
\end{equation}
therefore, we can switch the integration over $\lambda$ and $q$. The solid angle $\diff\Pi$ then reads

\begin{equation}
 \diff\Pi = \frac{1}{d^2_o}\left|\frac{\partial(\alpha,\beta)}{\partial(\lambda,q)}\right|\diff\lambda\diff q.
\end{equation}
where $|\partial{\alpha,\beta}/\partial{\lambda,q}|$ is the Jacobian of the transformation $(\alpha,\beta)\rightarrow (\lambda,q)$ and reads

\begin{equation}
 \left|\frac{\partial(\alpha,\beta)}{\partial(\lambda,q)}\right|=\frac{q}{\sin\theta_{o}\sqrt{q^2+a^2\cos^2\theta_{o}-\lambda^2\cot^2\theta_{o}}}.
\end{equation}
The impact parameters $\lambda$ and $q$ can be expressed in terms of the azimuthal, $\phi_e$ , and radial , $r_e$ , position of the emitter. 

We calculate derivatives $\partial r_e/\partial\lambda$ and $\partial r_e/\partial q$ from the condition 

\begin{eqnarray}
 H(r_e,\lambda,q)&\equiv& u_{sgn}\left[\int_{1/r_e}^{u_t}\frac{\diff u}{\sqrt{U}}+(-1)^{n_u-1}\int_{u_o}^{u_t}\frac{\diff u}{\sqrt{U}}\right]\nonumber\\
&-&\mu_{sgn}\left[\int_{\mu_e}^{\mu_\pm}\frac{\diff\mu}{\sqrt{M}}+(-1)^{n_\mu-1}\int_{\mu_o}^{\mu_\pm}\frac{\diff u}{\sqrt{U}}+p\int_{\mu_\mp}^{\mu_\pm}\frac{\diff\mu}{\sqrt{M}}\right]=0,
\end{eqnarray}
where the turning points  $u_t$ and $\mu_t$ are functions of the impact parameters $\lambda$ and $q$ as well as the radial and latitudinal motion functions $U=U(u,\lambda,q)$ and $M=M(\mu,\lambda,q)$. The functions $U$ and $M$ give the radial and latitudinal photon motion in terms of the variables $u\equiv 1/r$ and $\mu\equiv\cos\theta$ (see \cite{Rau-Bla:1994:}) by the relations
\begin{eqnarray}
 U(u;\lambda,q)&\equiv& 1 + (a^2-\lambda^2-q^2)u^2 + 2[(\lambda-a)^2+q^2]u^3\nonumber\\
&& - [b(a - \lambda)^2 + (a^2 + b)q^2]u^4,\\
M(\mu;\lambda,q)&\equiv& q^2 + (a^2-\lambda^2-q^2)\mu^2 - a^2\mu^4.
\end{eqnarray}
$H$ is an implicit function of $r_e$, $\lambda$ and $q$. From the implicit function theorem we find the derivatives to be

\begin{equation}
 \frac{\partial r_e}{\partial \lambda}=-\frac{\partial H/\partial\lambda}{\partial H/\partial r_e};\quad  \frac{\partial r_e}{\partial q}=-\frac{\partial H/\partial q}{\partial H/\partial r_e},
\end{equation}
and after introducing the parameter 

\begin{equation}
p=[n_\mu(1-\mathrm{mod}(n_\mu,2))+(n_\mu-1)\mathrm{mod}(n_\mu,2)] 
\end{equation}
where $n_u$($n_\mu$) determines the number of turning points of the photon radial (latitudinal) motion, we arrive to the relation

\begin{eqnarray}
 \frac{\partial H}{\partial s}&=&u_{sgn}\left[\int_{1/r_e}^{u_t}{-\frac{\partial U/\partial s}{2U^{3/2}}\diff u}+(-1)^{n_u-1}\int_{u_o}^{u_t}{-\frac{\partial U/\partial s}{2U^{3/2}}\diff u}\right]\nonumber\\
&+&u_{sgn}\frac{\partial u_t}{\partial s}\frac{1}{\sqrt{U(u_t)}}(1+(-1)^{n_u-1})\nonumber\\
&-&\mu_{sgn}\left[\int_{\mu_e}^{\mu_\pm}{-\frac{\partial M/\partial s}{2M^{3/2}}}+(-1)^{n_\mu-1}\int_{\mu_o}^{\mu_\pm}{-\frac{\partial M/\partial s}{2M^{3/2}}}+p\int_{\mu_\mp}^{\mu_\pm}{-\frac{\partial M/\partial s}{2M^{3/2}}}\right]\nonumber\\
&-&\mu_{sgn}\left[\frac{\partial\mu_\pm}{\partial s}\frac{1}{\sqrt{M(\mu_\pm)}}(1+(-1)^{n_\mu-1}+p)-\frac{p}{\sqrt{M(\mu_\mp)}}\frac{\partial\mu_\mp}{\partial s}\right]\label{dHds}
\end{eqnarray}
with $s$ being replaced for $\lambda$ or $q$, and

\begin{equation}
 \frac{\partial H}{\partial r_e}=\frac{1}{\sqrt{U(1/r_e)}r_e^2}.
\end{equation}

If $n_u=0$ and $n_\mu=0$ this formulas are easily enumerable, but in other cases there is a problem with integration in turning points (the integrals diverge) and there are some other terms which also diverge. Expressing the effective potentials in the form (see \cite{Viergutz:1993:})

\begin{equation}
 U(u,\lambda,q)\equiv(u_t-u)P(u,\lambda,q);\quad  M(\mu,\lambda,q)\equiv(\mu_\pm-\mu)W(\mu,\lambda,q) \label{UM}
\end{equation}
we can avoid this problem. As an example let's take $n_u=1$ and transform the term

\begin{equation}
T=\int_{1/r_e}^{u_t}{-\frac{\partial U/\partial s}{2U^{3/2}}\diff u}+\frac{\partial u_t}{\partial s}\frac{1}{\sqrt{U(u_t)}}
\end{equation}
using (\ref{UM}). After some algebra one arrives to formula

\begin{equation}
 T=\int_{1/r_e}^{u_t}{-\frac{(\partial P/\partial s) + (\partial P/\partial u)(\partial u_t/\partial s)}{2P\sqrt{U}}}+\frac{\partial u_t}{\partial s}\left(\frac{1}{\sqrt{u_e}}+\frac{1}{\sqrt{u_o}}\right),
\end{equation}
which is now enumerable. The terms in $\mu$ part of (\ref{dHds}) are transformed in the same way. The partial derivatives of the turning points can be also found from the implicit formula

\begin{equation}
 U(u_t,\lambda,q)=0\quad\Rightarrow\quad \frac{\partial u_t}{\partial s}=-\frac{\partial U/\partial s}{\partial U/\partial u_t},
\end{equation}
where, again, $s$ is replaced with $\lambda$ and $q$. 

From the Carter equation for the azimuthal coordinate 

\begin{eqnarray}
 \phi_e &=& -u_{sgn}\left[\int_{1/r_e}^{u_t}{\frac{f_U}{\sqrt{U}}\diff u}+(-1)^{n_u-1}\int_{u_o}^{u_t}{\frac{f_U}{\sqrt{U}}\diff u}\right]\\ 
&-&\mu_{sgn}\left[\int_{\mu_e}^{\mu_\pm}{\frac{f_M}{\sqrt{M}}\diff \mu}+(-1)^{n_\mu-1}\int_{\mu_o}^{\mu_\pm}{\frac{f_M}{\sqrt{M}}+p\int_{\mu_\mp}^{\mu_\pm}{\frac{f_M}{\sqrt{M}}\diff \mu}}\right] 
\end{eqnarray}
with

\begin{equation}
f_U=[2(a-l)u+l]/(1-2u+a^2 u^2) 
\end{equation}
and
\begin{equation}
f_M=l\mu^2/(1-\mu^2),
\end{equation}
we calculate the derivatives $\partial\phi_e/\partial\lambda$ and $\partial\phi_e/\partial q$ having in mind that $r_e=r_e(\lambda,q)$. From the above calculations we can form the Jacobian of transformation  $(\lambda,q) \rightarrow (r_e,\phi_e)$ in the form

\begin{equation}
 \left|\frac{\partial(\lambda,q)}{\partial(r_e,\phi_e)}\right|=\left|\frac{\partial(r_e,\phi_e)}{\partial(\lambda,q)}\right|^{-1}=\left|\frac{\partial r_e}{\partial\lambda}\frac{\partial \phi_e}{\partial q}-\frac{\partial r_e}{\partial q}\frac{\partial \phi_e}{\partial\lambda}\right|^{-1}.
\end{equation}
The formula for the solid angle $\diff\Pi$ then reads

\begin{equation}
 \diff\Pi =\frac{q}{\sin\theta_o\sqrt{q^2+a^2\cos^2\theta_o-\lambda^2\cot^2\theta_o}}\left|\frac{\partial r_e}{\partial\lambda}\frac{\partial \phi_e}{\partial q}-\frac{\partial r_e}{\partial q}\frac{\partial \phi_e}{\partial\lambda}\right|^{-1}\diff r_e\diff\phi_e.\label{SolidAngle}
\end{equation}

\section{Light curve of a hot spot on a circular keplerian orbit}
We model following situation. A monochromatic isotropically radiating hot spot follows the circular geodesic with radius $r_e$ in the equatorial plane. The flux that observer at infinity measures is

\begin{equation}
 \diff F_o = I_o \diff\Pi, \label{LightCurveFlux}
\end{equation}
 where $I_o=g^4 I_e$ \cite{MTW}. $I_e$ is a function of radial position of the emitter. Since we study here the light curve of a single hot spot on the circular geodesic, we normalize $I_e =1$.
Putting (\ref{SolidAngle}) into (\ref{LightCurveFlux}), the differential of the observed flux then reads 

\begin{equation}
 \diff F_o =\frac{1}{d_o^2}\frac{q g^4(\phi_e,r_e)}{\sin\theta_o\sqrt{q^2+a^2\cos^2\theta_o-\lambda^2\cot^2\theta_o}}\left|\frac{\partial r_e}{\partial\lambda}\frac{\partial \phi_e}{\partial q}-\frac{\partial r_e}{\partial q}\frac{\partial \phi_e}{\partial\lambda}\right|^{-1}\diff r_e\diff\phi_e.
\end{equation}
The light curve of such a radiating spot could be obtained by introducing the time dependence of the radiation received by the distant observer, including the time-delay effects related to the relativistic motion \cite{Bao-Stu:1992:}. Here we consider another situation, namely the stationary line profile generated by a ring of orbiting hot spots, postponing the light curves of isolated hot spots to future studies.

These results could be directly applied to calculate line profile generated by some part of the internal region of the accretion disc, assuming some reasonable emissivity law, since such a part of the disc can be represented as being composed from radiating rings. Such a study will be presented in the future study, now being under preparation.

\section{Profiled spectral lines}
We directly apply the results of the previous section to determine the spectral line profile of the bright ring in the equatorial plane of braneworld Kerr black hole.

Let the source radiates isotropically at a fixed frequency $\nu_{e}$. The specific intensity $I_{e}$ of the source is then given by

\begin{equation}
 I_{e}(\nu_{e})=\epsilon(r)\delta(\nu_{e}-\nu_{0}),\label{intensity}
\end{equation}
where $\epsilon(r)$ is the local emissivity given as a function of the radial coordinate of the black hole spacetime, $\nu_0$ is the rest frequency. Using (\ref{intensity}) and (\ref{flux}) we arrive to the formula for the specific flux in the form

\begin{equation}
 F_{o}(\nu_{o}) = \int\epsilon(r)g^4\delta(\nu_{o}-g\nu_{0})\diff\Pi.\label{Flux0}
\end{equation}
Here, since we calculate the profiled spectral line of a single radiating ring, the emissivity function can be normalized by $\epsilon(r)=1$. We rearrange the  solid angle formula (\ref{SolidAngle}) for the calculation of the specific flux to the form

\begin{equation}
 \diff\Pi = \frac{q}{\sin\theta_o\sqrt{q^2+a^2\cos^2\theta_o-\lambda^2\cot^2\theta_o}}\left|\frac{\partial r_e}{\partial\lambda}\frac{\partial g}{\partial q}-\frac{\partial r_e}{\partial q}\frac{\partial g}{\partial\lambda}\right|^{-1}\diff g\diff r_e\label{SolidAngleSpec}.
\end{equation}
Using Eqs. (\ref{SolidAngleSpec}) and (\ref{Flux0}) we finally arrive to

\begin{equation}
 F_{o}(g) = \int{\frac{q}{d_o^2\sin^2\theta_o\sqrt{q^2+a^2\cos^2\theta_o-\lambda^2\cot^2\theta_o}}}\left|\frac{\partial r_e}{\partial\lambda}\frac{\partial g}{\partial q}-\frac{\partial r_e}{\partial q}\frac{\partial g}{\partial\lambda}\right|^{-1}\diff r_e.\label{flux_tot}
\end{equation}
In order to obtain the spectral line profile form (\ref{flux_tot}), one must find all relevant doubles of ($\lambda$,$q$), which are related to doubles of $g$ and $\chi$, by integrating  the Carter equations schematically given by the formula

\begin{equation}
 \pm\int_{\theta e}\frac{1}{\sqrt{W(\theta)}}\diff\theta =\pm\int_{re}\frac{1}{\sqrt{R(r)}}\diff r 
\end{equation}
The detailed calculations in terms of the radial and latitudinal variables $u$ and $\mu$ and the motion functions $U(u,\lambda,q)$, $M(\mu,\lambda,q)$ are done in terms of the elliptic integrals that are expressed using the following tables of integrals \cite{SS:a:RAGTime:2007:Proceedings}. 

\begin{table}[!th]
\caption{The reductions of $\int^m_{m_1}\diff m'/\sqrt{M(m')}=I_M$}

\begin{tabular}{lllll}
\hline
  	Case & $\tan\Psi$ & $m$ & $c_1$ & $m_1$\\
	\hline\\
	$M_-<0$ & $\sqrt{\frac{M_+}{m^2}-1}$ & $\frac{M_+}{M_+-M_-}$ & $\frac{1}{\sqrt{a^2(M_+-M_-)}}$ & $\sqrt{M_+}$\\
	\\
	$M_->0$ & $\sqrt{\frac{M_+-m^2}{m^2-M_-}}$ & $\frac{M_+-M_-}{M_+}$ &
        $\frac{1}{a^2}$ & $\sqrt{M_+}$\\
        \hline
 \end{tabular}\label{tableEIM}

\end{table} 

\begin{table}[!th]
\caption{The reductions of $\int^u_{u_1}\diff u'/\sqrt{U(u')}=I_U$}

\begin{tabular}{lllll}
\hline
  	Case & $\tan\Psi$ & $m$ & $c_1$ & $u_1$\\
	\hline
	I & $\sqrt{\frac{(\beta_1-\beta_3)(u-\beta_4)}{(\beta_1-\beta_4)(\beta_3-u)}}$ & $\frac{(\beta_1-\beta_2)(\beta_3-\beta_4)}{(\beta_1-\beta_3)(\beta_2-\beta_4)}$ & $\frac{2}{\sqrt{\tilde{q}(b1-b3)(b2-b4)}}$ & $\beta_4$\\
	\\
	II & $\sqrt{\frac{(\beta_1-\beta_2)(u-\beta_3)}{(\beta_1-\beta_3)(\beta_2-u)}}$ & $\frac{(\beta_2-\beta_3)(\beta_1-\beta_4)}{(\beta_1-\beta_2)(\beta_4-\beta_3)}$ & $\frac{2}{\sqrt{-\tilde{q}(b1-b2)(b3-b4)]}}$ & $\beta_3$\\
	\\
	III & $\frac{2c_2(u)}{|1-c^2_2(u)|}$ & $\frac{4c_4 c_5 - (\beta_3 - \beta_4)^2 - c_4 c_5}{4c_4 c_5}$ & $\frac{1}{\sqrt{-\tilde{q}c_4 c_5}}$ & $\beta_3$\\
	\\
	IV & $\frac{u-c_3}{\Im(\beta_1)(1+c_2^2)+c_2(u-c_3)}$ & $1-\left(\frac{c_4-c_5}{c_4+c_5}\right)^2$ & $\frac{2}{(c_4+c_5)\sqrt{-\tilde{q}}}$ & $c_3$\\
	\\
	V & $\frac{2c_2(u)}{|1-c^2_2(u)|}$ & $1-\frac{(c_4+c_5)^2-(\beta_1 -
          \beta_4)^2}{4c_4 c_5}$ & $\frac{1}{\sqrt{\tilde{q}c_4 c_5}}$ &
        $\beta_4$\\
        \hline
 \end{tabular}\label{tableEI}

\end{table}

\begin{table}[!th]
\caption{Definitions for Table \ref{tableEI}.}\label{tableEI2}
\begin{tabular}{lll}
  \hline
  	Case & $^1 c_2$ & $^1 c_3$\\
	\hline
	III &  $\sqrt{\frac{c5(u-\beta_3)}{c_4(u-\beta_4)}} $ & -\\
	\\
	IV & $\sqrt{\frac{4[\Im(\beta_1)]^2-(c_4-c_5)^2}{(c_4+c_5)^2-4[\Im(\beta_1)]^2}}$ & $\mbox{\fontsize{8}{10}\selectfont $\Re(\beta_1)+c_2\Im(\beta_1)$}$\\
	\\
	V & $\sqrt{\frac{c4(u-\beta_4)}{c_5(\beta_1-u)}} $& -\\
        \hline
\end{tabular}
\end{table}

\begin{table}[!th]
\caption{Definitions for Table \ref{tableEI} and Table \ref{tableEI2}.}

\begin{tabular}{lll}
  \hline
  	Case & $^1 c_4$ & $^1 c_5$\\
	\hline
	III &   $\mbox{\fontsize{8}{10}\selectfont$\sqrt{\left[\Re(\beta_1)-\beta_3\right]^2+[\Im(\beta_1)]^2}$}$ & $\mbox{\fontsize{8}{10}\selectfont $\sqrt{\left[\Re(\beta_1)-\beta_4\right]^2+[\Im(\beta_1)]^2}$}$\\
	\\
	IV & $\mbox{\fontsize{8}{10}\selectfont $\sqrt{\left[\Re(\beta_1)-\Re(\beta_3)\right]^2+[\Im(\beta_1)+\Im(\beta_3)]^2}$}$ & $\mbox{\fontsize{8}{10}\selectfont $\sqrt{\left[\Re(\beta_1)-\Re(\beta_3)\right]^2+[\Im(\beta_1)-\Im(\beta_3)]^2}$}$\\
	\\
	V & $\mbox{\fontsize{8}{10}\selectfont $\sqrt{\left[\Re(\beta_2)-\beta_1\right]^2+[\Im(\beta_2)]^2}$}$ & $\mbox{\fontsize{8}{10}\selectfont $\sqrt{\left[\Re(\beta_2)-\beta_4\right]^2+[\Im(\beta_2)]^2}$}$\\
	\hline
\end{tabular}
\begin{tabular}{c}
	$^1$\textit{The symbols $\Re(x)$ and $\Im(x)$  refer to real and imaginary part of $x$ here.}
\end{tabular}

\end{table}

We express the radial and latitudinal motion integrals in the form of the standard elliptic integrals of the first kind. Rauch and Blandford presented the tables of reductions of $u$-integrals and $\mu$-integrals for the case of photons in Kerr geometry \cite{Rau-Bla:1994:}. Here we extended those reductions for the case of nonzero braneworld parameter $b$. Because the integration of the $\mu$-integral does not depend on braneworld parameter $b$, the transformations are the same as in the case of Kerr metric (see \cite{Rau-Bla:1994:}). 

There are two cases we distinguish in the latitudinal motion integral (see table \ref{tableEIM}). In the first case there is one positive, $M_+>0$, and one negative, $M_-<0$ root of $M(\mu^2)$ and it implies two turning points located symmetrically about the equatorial plane, given by $\pm\sqrt{M_+}$ that corresponds to the so called orbital motion crossing the equatorial plane \cite{BS:1976:,CHANDRA}. In the second case, there are two positive roots, $0<M_-<M_+$ of  $M(\mu^2)$ implying that the latitudinal motion is constrained to the region above or below of the equatorial plane that correspond to the so called vortical motion \cite{CHANDRA}. The relevant reductions of integral 
$\int^\mu_{\mu_1}\diff \mu'/\sqrt{M(\mu')}=I_M$ are stored in the table \ref{tableEIM}.

 For distant observers we distinguish five relevant cases of the radial integral. These cases depend on the character of roots of the quartic equation $U(u)=0$, i.e., on the number of turning points ($n_u=0$ or $n_u=1$) in the radial motion  and the value of parameter $\tilde{q}=q(a^2+b)+b(a-l)^2$. Those transformations are presented in the table \ref{tableEI}.

Denoting roots of the quartic equation $U(u)=0$ by $\beta_1$, $\beta_2$, $\beta_3$ and $\beta_4$, the meaning of each of the five cases is the following:
\begin{itemize}
\item
The \textbf{case I}: four distinct real roots of $U(u)=0$ forming the sequence $\beta_1>\beta_2>\beta_2>0$ and $\beta_4<0$. The value of modified constant of motion $\tilde{q}>0$.
\item
The \textbf{case II}: four real roots as in the case I but their values form the following order: $\beta_1>\beta_2>0$ and $\beta_4<\beta_3<0$. The value of modified constant of motion  $\tilde{q}<0$. 
\item
The \textbf{case III}: two real and two complex roots of $U(u)=0$: $\beta_1$ being a complex root, $\beta_2=\bar{\beta_1}$ and $\beta_4<\beta_3<0$. The value of modified constant of motion $\tilde{q}<0$.
\item
The \textbf{case IV}: only complex roots: $\beta_2=\bar{\beta_1}$ and $\beta_4=\bar{\beta_3}$. The value of modified constant of motion $\tilde{q}<0$. 
\item
The \textbf{case V}: two real and two complex roots of $U(u)=0$: $\beta_1>0$, $\beta_4<0$, $\beta_2$ being a complex root and $\beta_3=\bar{\beta_2}$. 
\end{itemize}

The presented transformations enable us to write the motion integrals in the form

\begin{equation}
 \int^{u}_{u_1}\frac{1}{\sqrt{U(\tilde{u})}}\diff \tilde{u} = c_1\mathcal{F}(\Psi;m),\label{ellint}
\end{equation}
and 
\begin{equation}
 \int^{\mu}_{\mu_1}\frac{1}{\sqrt{M(\tilde{\mu})}}\diff \tilde{\mu} = c_1\mathcal{F}(\Psi;m),\label{ellintM}
\end{equation}
where $\mathcal{F}$ is the elliptic integral of the first kind and $u_1$(resp $\mu_1$) depends on the case of root distribution of quartic equation $U(u)=0$ ($M(\mu)=0$) as given in Table \ref{tableEI} (\ref{tableEIM}). If, in the cases III and V, there is the value of $1-c_2^2(u)<0$, we have to take instead of  (\ref{ellint}) the integral in the form

\begin{equation}
 \int^{u}_{u_1}\frac{1}{\sqrt{U(\tilde{u})}}\diff \tilde{u} =  c_1(2\mathcal{K}(m)-\mathcal{F}(\Psi;m)),\label{ellint1}
\end{equation}
where $\mathcal{K}$ is the complete elliptic integral of the first kind. In the case that $sign(\mu_1\cdot \mu)<0$ we have to take instead of (\ref{ellintM}) the integral in the form
\begin{equation}
 \int^{\mu}_{\mu_1}\frac{1}{\sqrt{M(\tilde{\mu})}}\diff \tilde{\mu} =  c_1(2\mathcal{K}(m)-\mathcal{F}(\Psi;m)),\label{ellintM1}
\end{equation}
where $\Psi$, $m$ and $c_1$ are taken from table \ref{tableEIM} and $\mathcal{K}$ is the complete elliptic integral of the first type.
 We consider two basic possibilities of trajectories, namely those corresponding to direct and indirect images.



\section{Results and discussion}
We have demonstrated the influence of the braneworld tidal charge on the profiled lines using a computational code developed under all the phenomena discussed above. Both the direct and indirect photons were considered. The profiled lines depend on the black hole parameters $a$ and $b$, on the radius $r_e$ of the radiating ring and on the inclination angle $\theta_0$ of the observer. We separate the results into two groups according to the sign of the tidal charge; we would like to stress that the results obtained for $b>0$ are relevant also for radiating rings of uncharged matter orbiting Kerr-Newman black holes with $b\rightarrow Q^2$ where $Q$ is the black hole charge parameter.

\begin{figure}[ht]
	\begin{tabular}{ccc}
		\includegraphics[width=3.5cm]{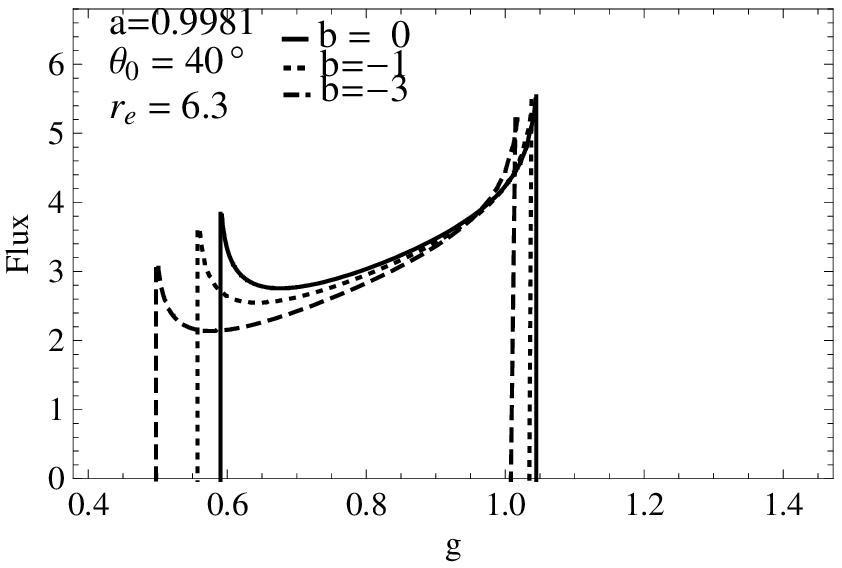}&\includegraphics[width=3.5cm]{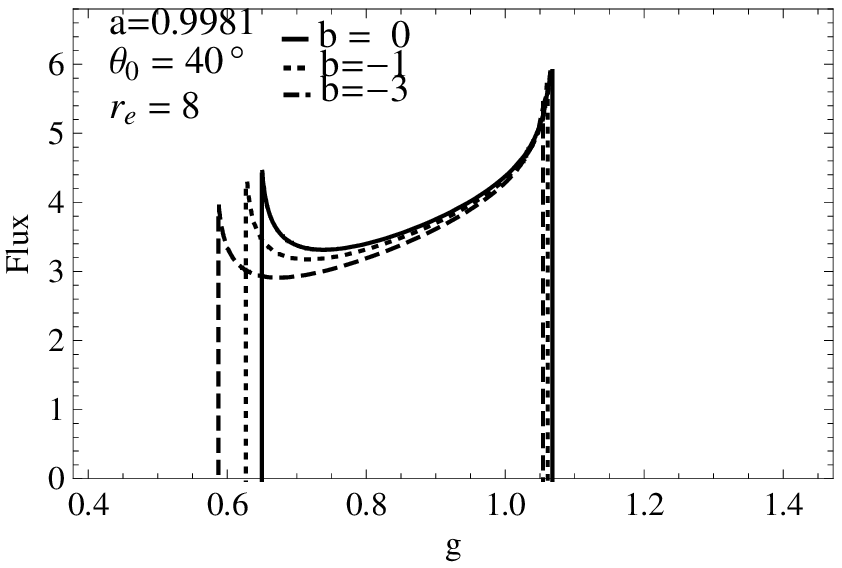}&\includegraphics[width=3.5cm]{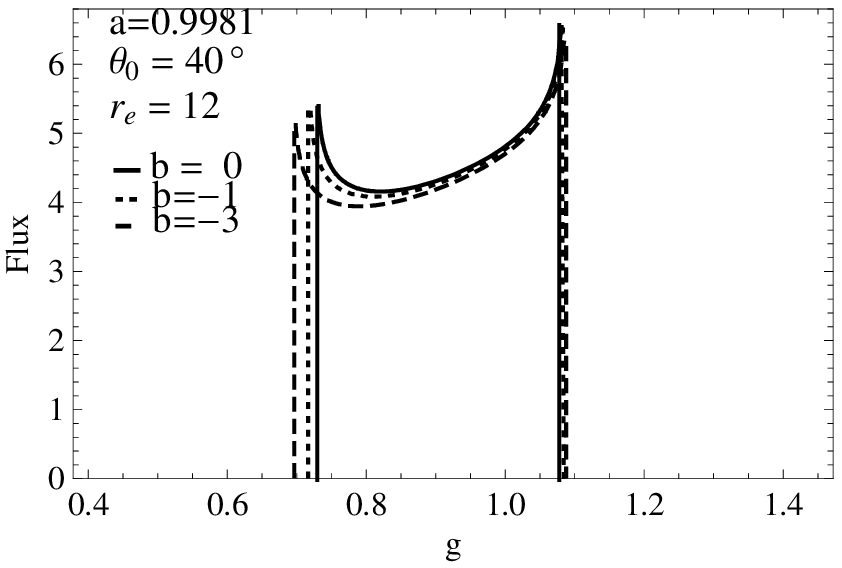}\\
		\includegraphics[width=3.5cm]{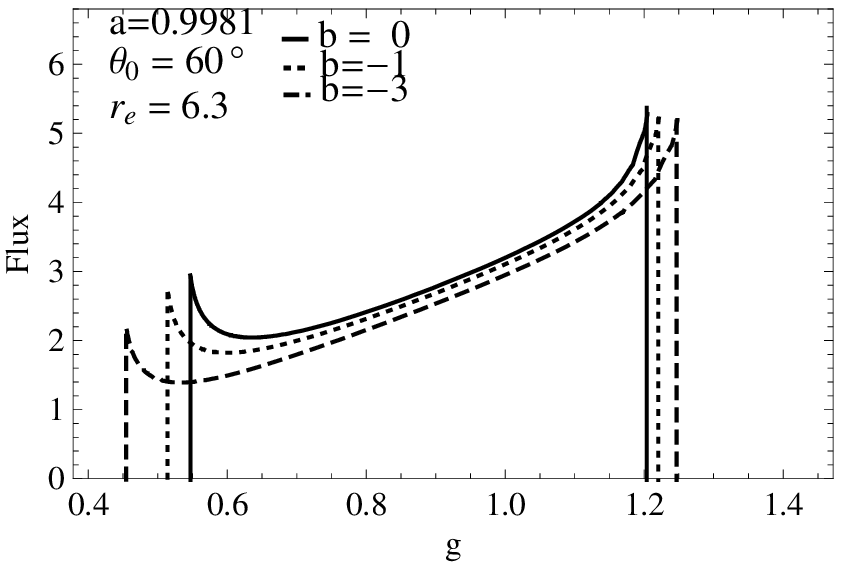}&\includegraphics[width=3.5cm]{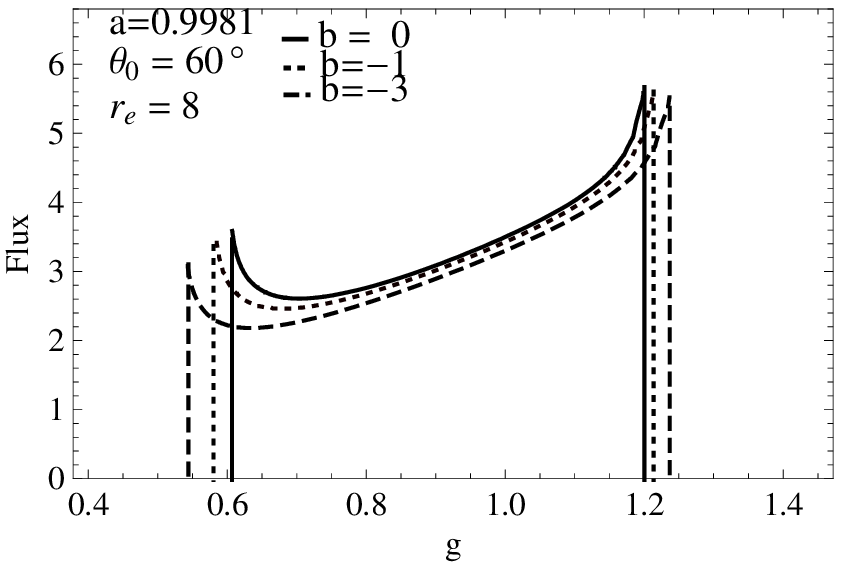}&\includegraphics[width=3.5cm]{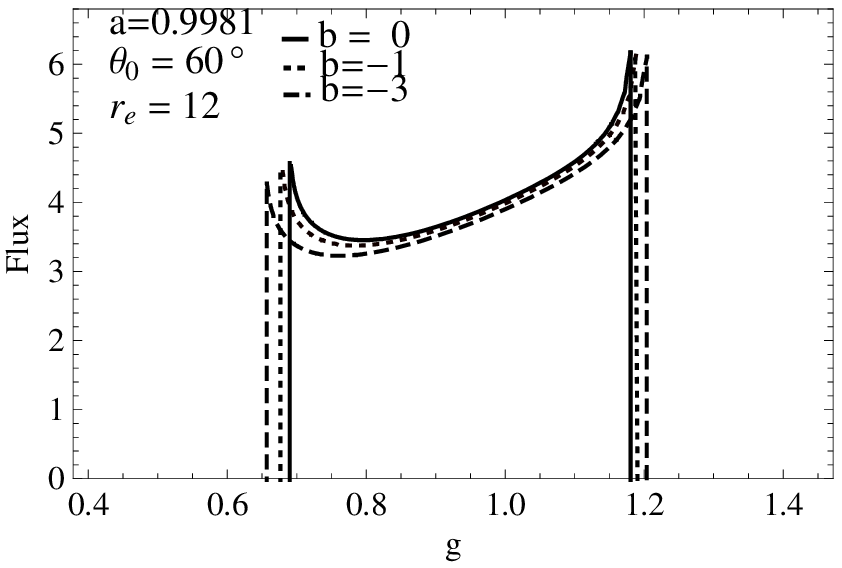}\\
		\includegraphics[width=3.5cm]{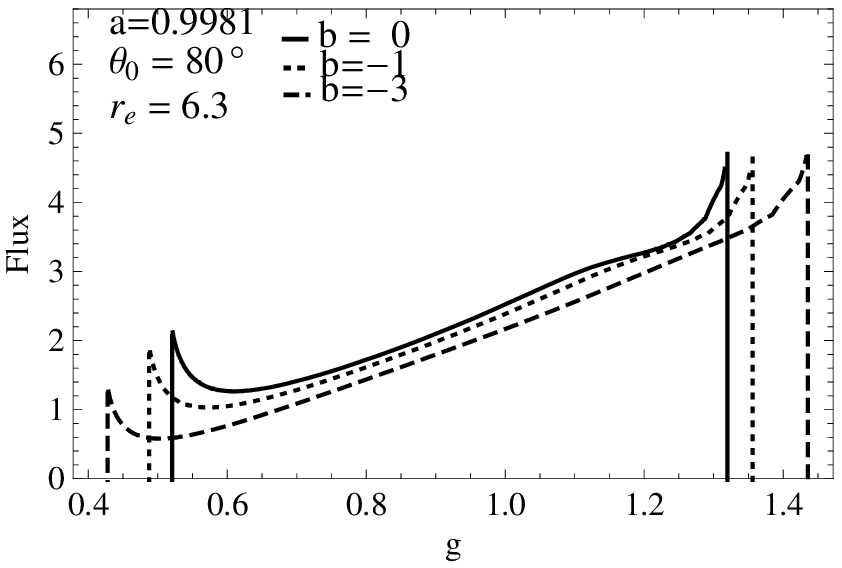}&\includegraphics[width=3.5cm]{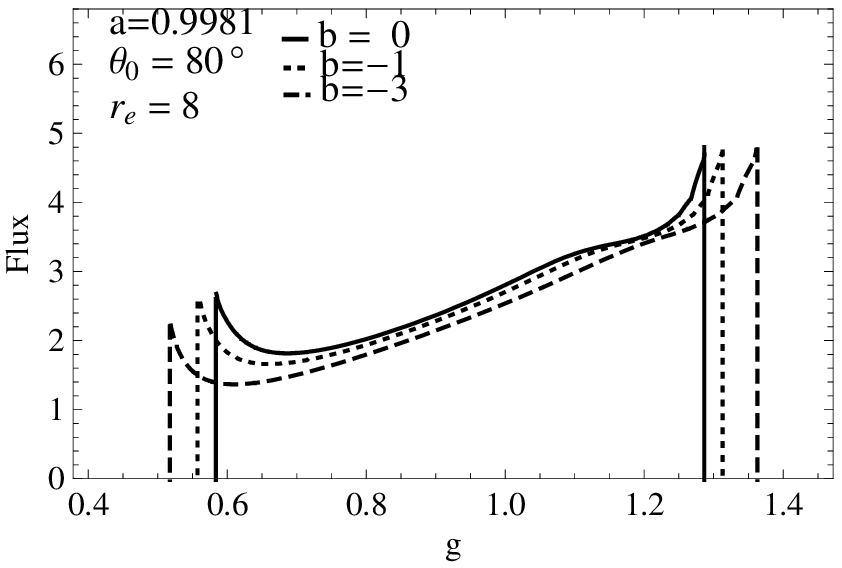}&\includegraphics[width=3.5cm]{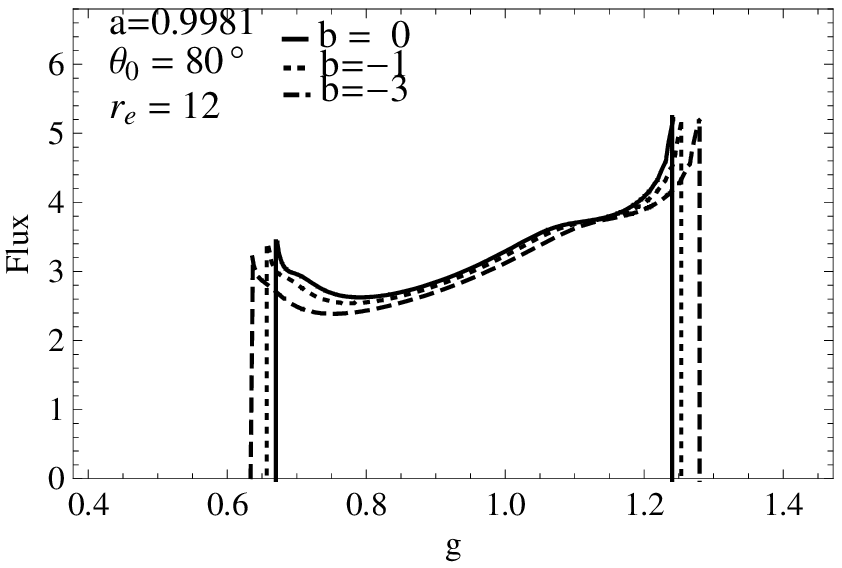}
	\end{tabular}
\caption{Line profiles for braneworld black holes with $b<0$. We demonstrate the influence of inclination angle of observer $\theta_0$, radial distance of emitter $r_e$ and braneworld parameter $b$ on the profile of spectral line of radiation emitter from radiating keplerian ring. With rotation parameter fixed to $a=0.9981$ each figure contains three plots for three representative values of braneworld parameter $b=-3$, $-1$ and $0$. The inclination angle $\theta_0$ and the ring (Boyer - Lindquist) radius $r_e$ are given in the figure.}\label{figure1}
\end{figure}

\begin{figure}[th]
	\begin{tabular}{ccc}
		\includegraphics[width=3.5cm]{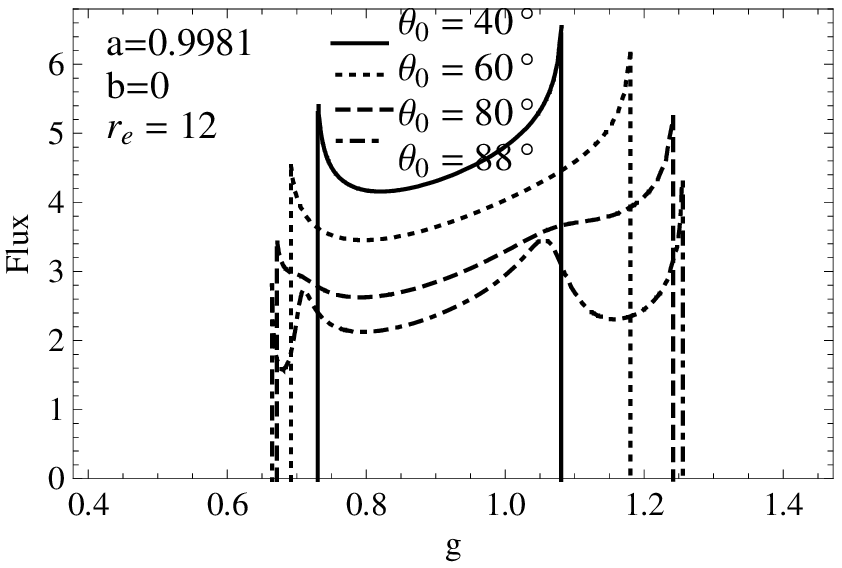}&\includegraphics[width=3.5cm]{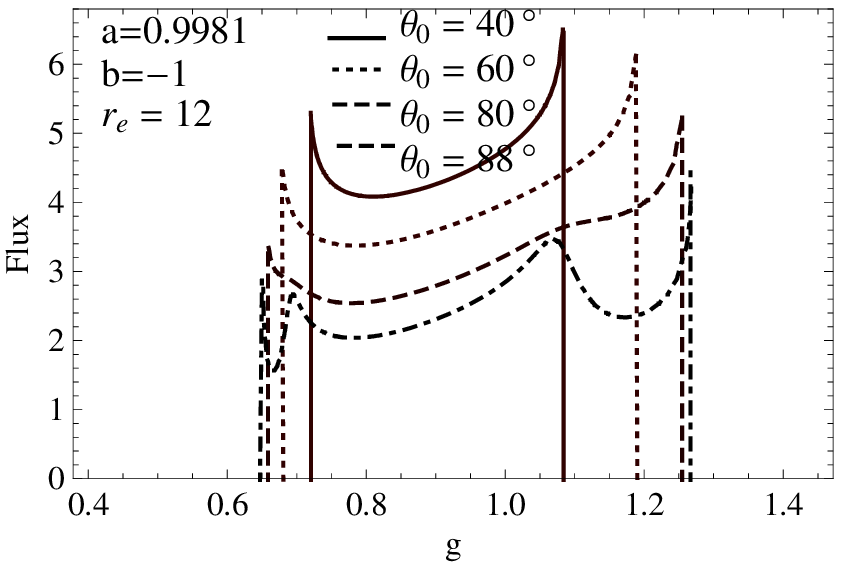}&\includegraphics[width=3.5cm]{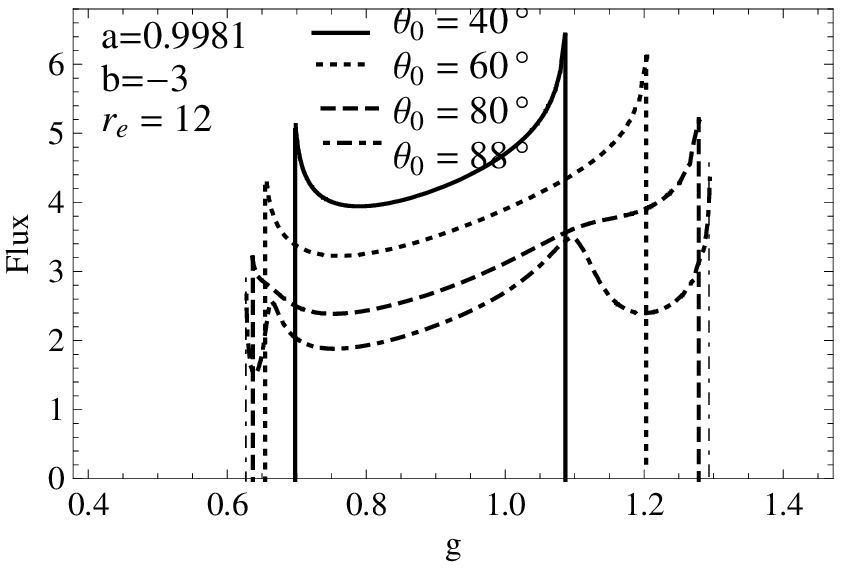}\\
		\includegraphics[width=3.5cm]{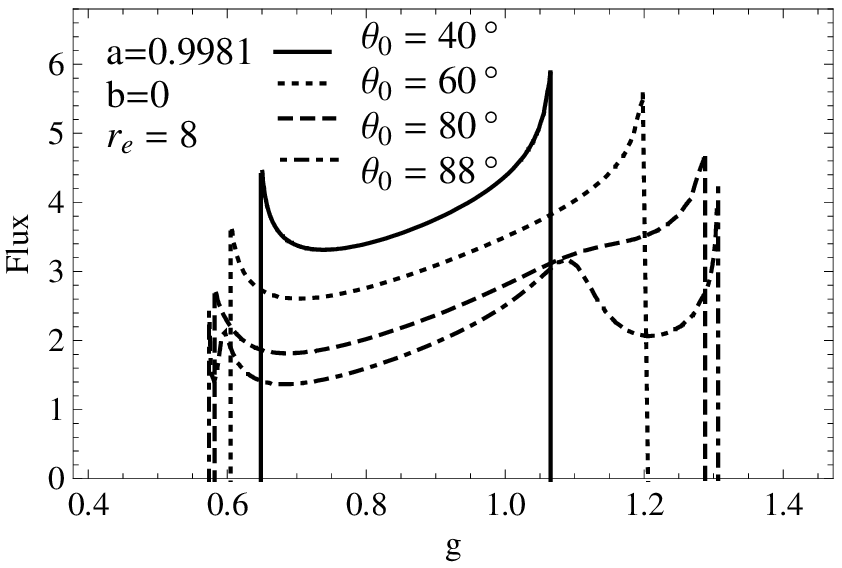}&\includegraphics[width=3.5cm]{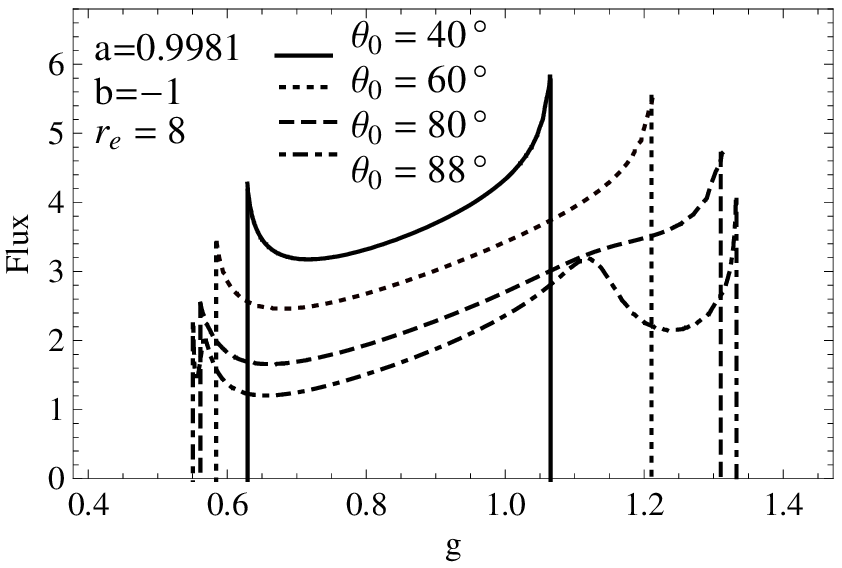}&\includegraphics[width=3.5cm]{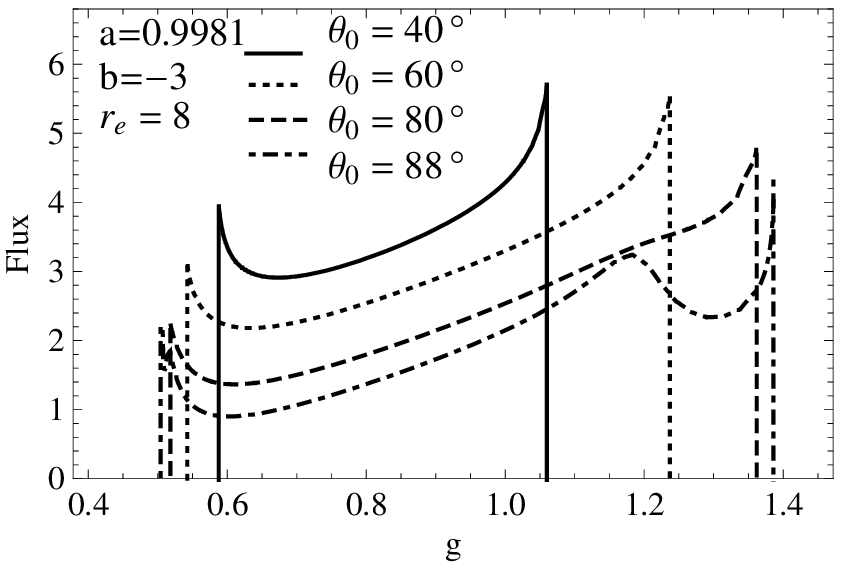}\\
		\includegraphics[width=3.5cm]{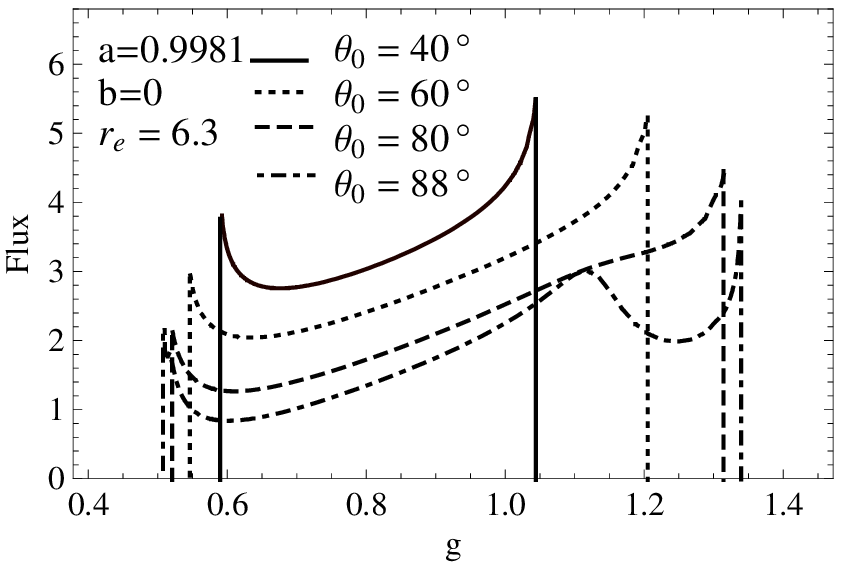}&\includegraphics[width=3.5cm]{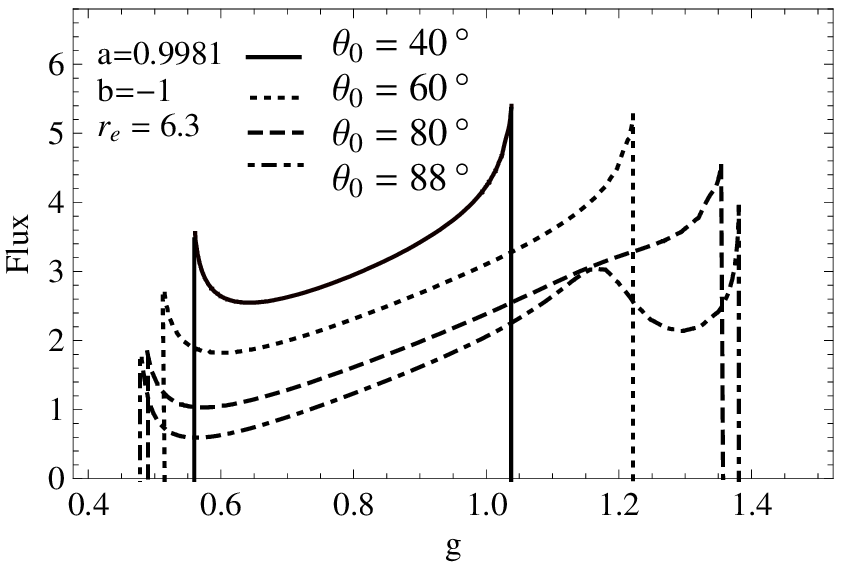}&\includegraphics[width=3.5cm]{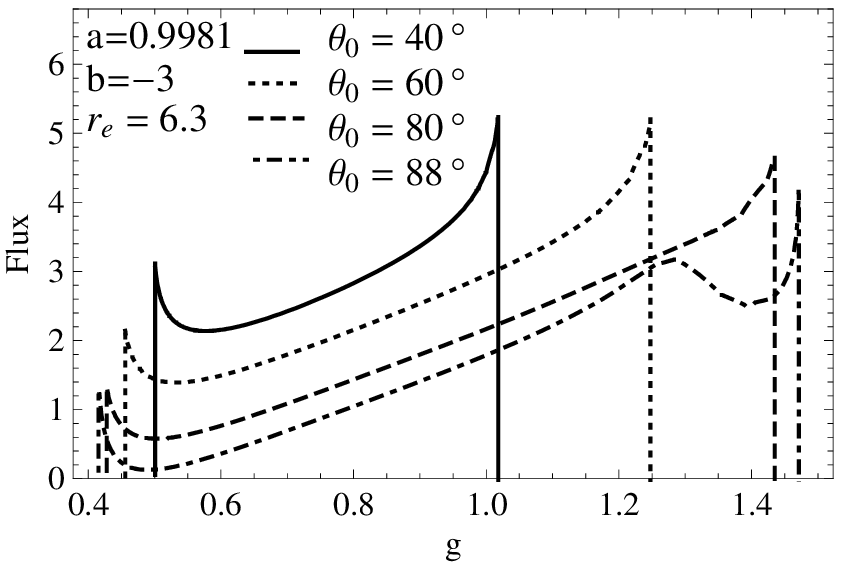}
	\end{tabular}
\caption{Line profiles for braneworld black holes with $b<0$. Each figure contain three plots for three representative values of $\theta_0=40^\circ$, $60^\circ$, $80^\circ$ and $88^\circ$. Figures in each row are plotted for fixed value of $r_e=12$(top), $8$(middle) and $6.3$(bottom) and braneworld parameter varying from left figure to right one with values $b=0$, $-1$ and $-3$.}\label{figure2}
\end{figure}

\begin{figure}[th]
	\begin{tabular}{ccc}
	\includegraphics[width=3.5cm]{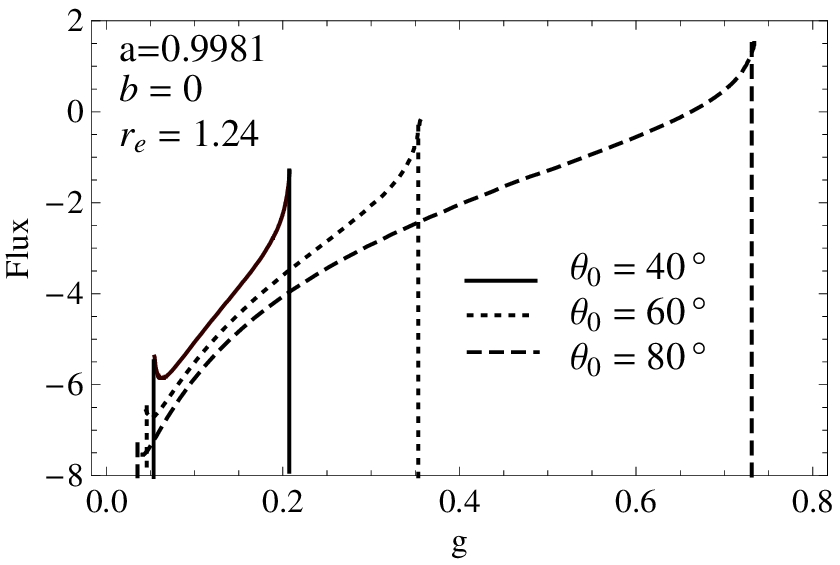}&\includegraphics[width=3.5cm]{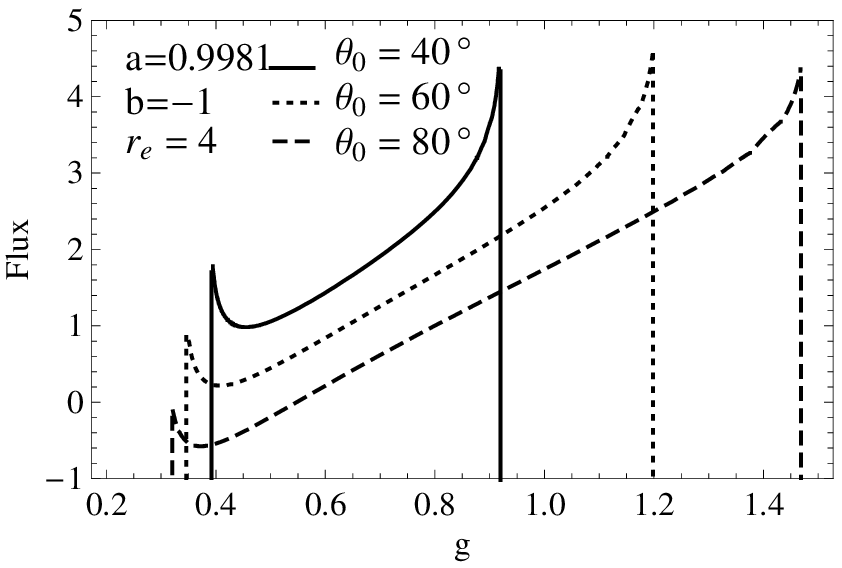}&\includegraphics[width=3.5cm]{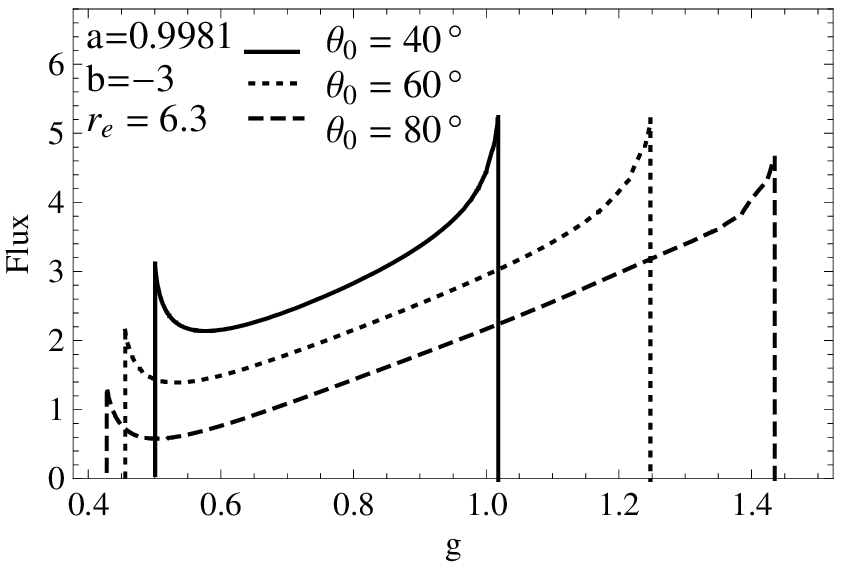}
	\end{tabular}
\caption{Line profiles for braneworld black holes with $b<0$. Each figure contains three plots of profiles of spectral line for three representative values of $\theta_0=40^\circ$, $60^\circ$ and $80^\circ$ and fixed(different for each figure) values of ($b=0$, $r_e=1.24$)(left), ($b=-1$, $r_e=4$)(middle) and ($b=-3$, $r_e=6.3$)(right). The emitters are on orbits close to marginally stable, which is function of braneworld parameter $b$, $r_{ms}=r_{ms}(b)$.}\label{figure3}
\end{figure}

\begin{figure}[th]
	\begin{tabular}{ccc}
		\includegraphics[width=3.5cm]{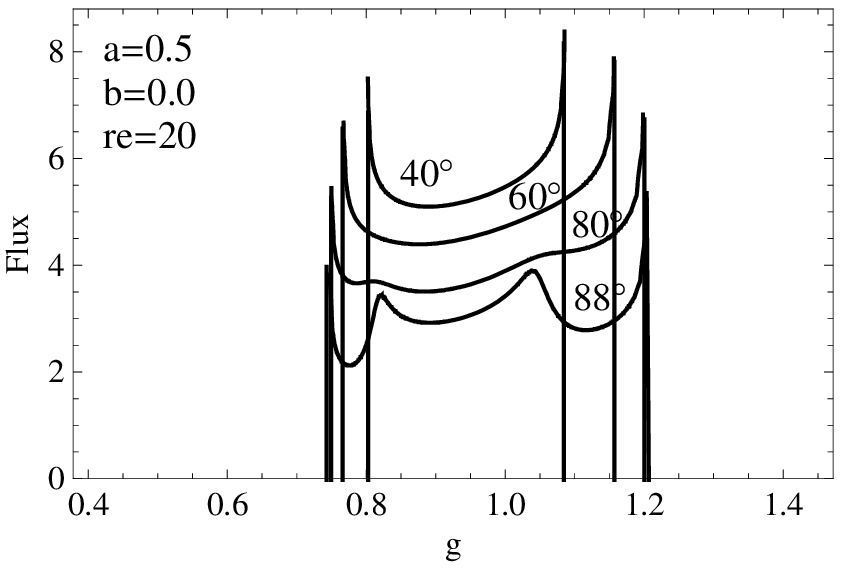}&\includegraphics[width=3.5cm]{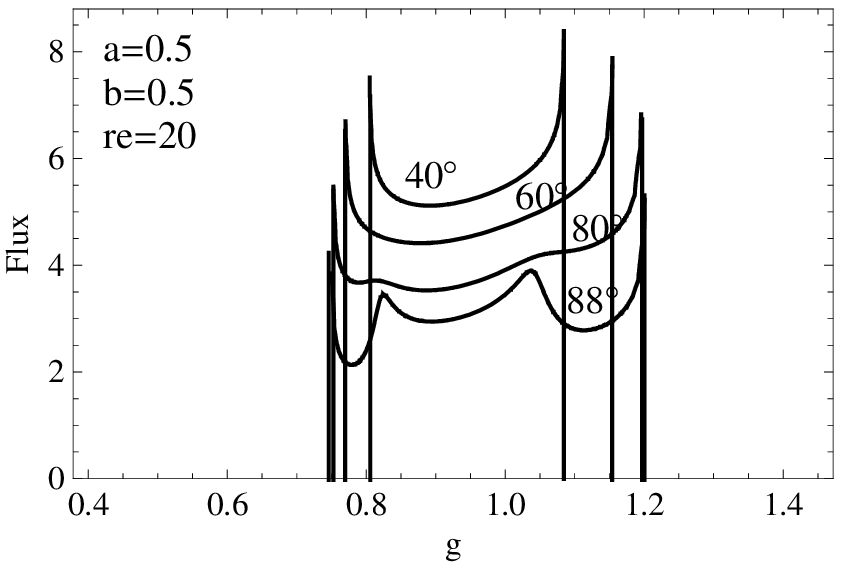}&\includegraphics[width=3.5cm]{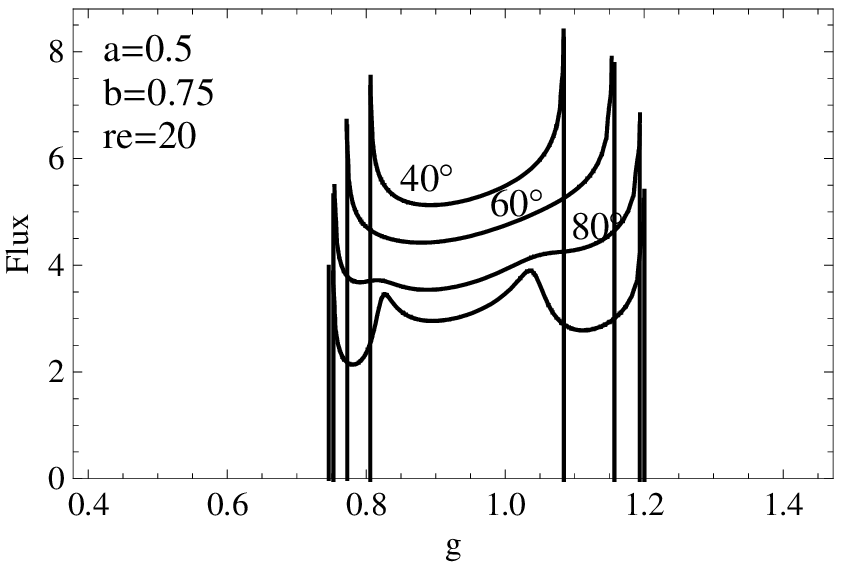}\\
		\includegraphics[width=3.5cm]{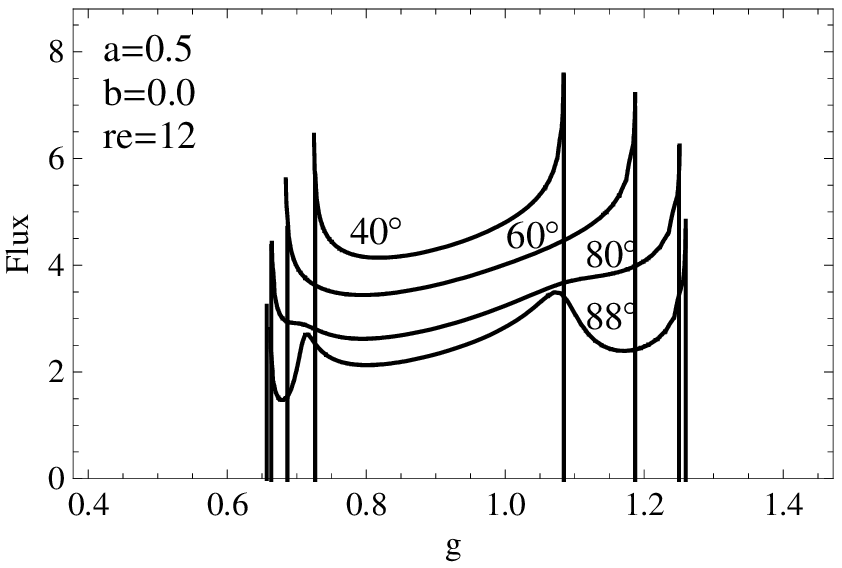}&\includegraphics[width=3.5cm]{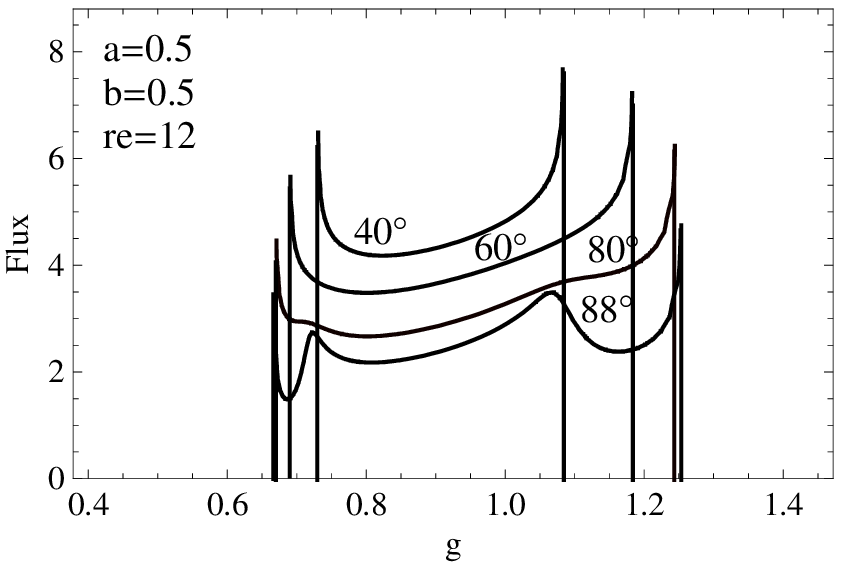}&\includegraphics[width=3.5cm]{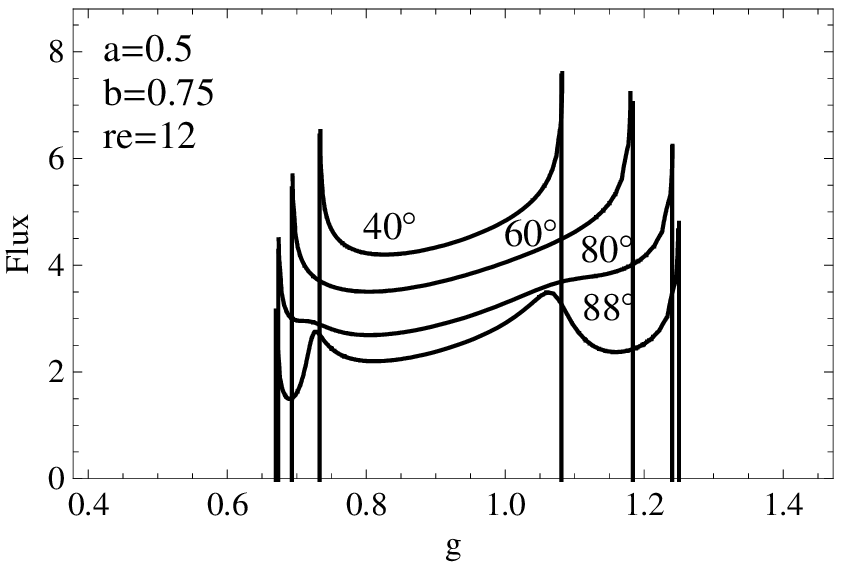}\\
		\includegraphics[width=3.5cm]{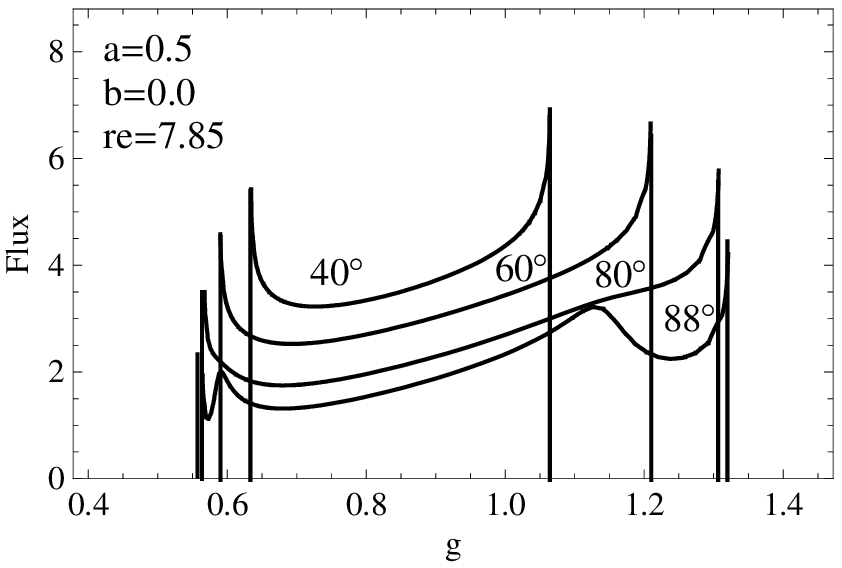}&\includegraphics[width=3.5cm]{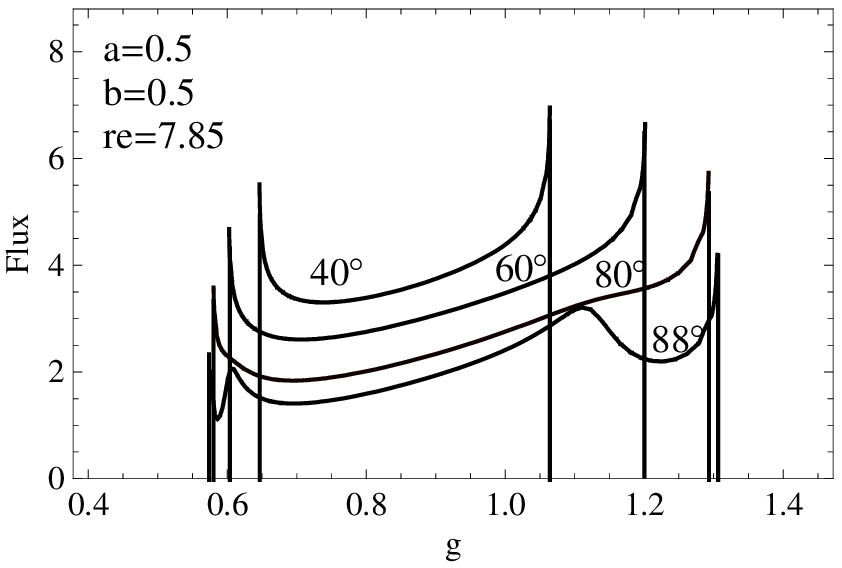}&\includegraphics[width=3.5cm]{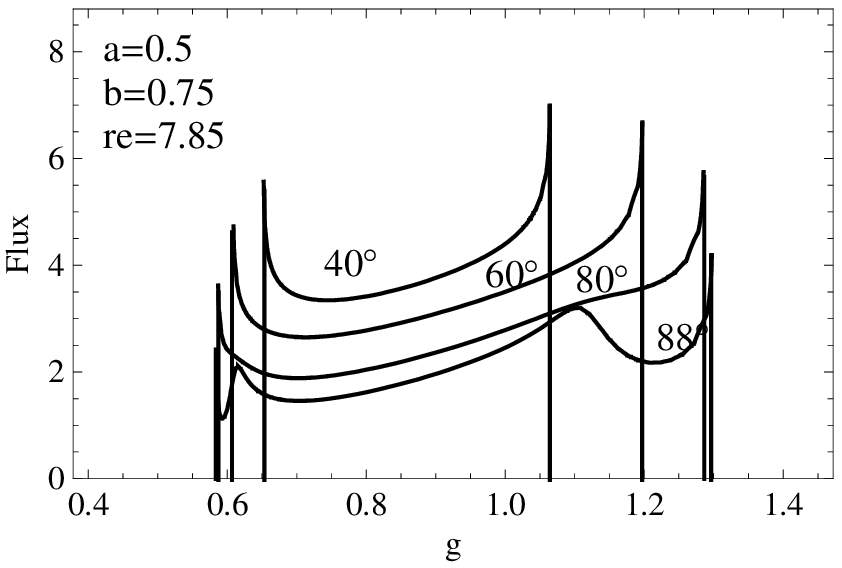}
	\end{tabular}
\caption{Line profiles for braneworld black holes with $b>0$. Each figure contain four plots for four representative values of $\theta_0=40^\circ$, $60^\circ$, $80^\circ$ and $88^\circ$. Figures in each row are plotted for fixed value of $r_e=20$(top), $12$(middle) and $7.85$(bottom) and braneworld parameter varying from left figure to right one with values $b=0$, $0.5$ and $0.75$.}\label{figure4}
\end{figure}

\begin{itemize}
 \item $b<0$
For negative tidal charges we have fixed the value of the spin parameter to value $a=0.9981$ close to extreme black hole state ($a=1$) for Kerr spacetimes ($b=0$). We then illustrate the influence of the tidal charge $b<0$ for a relatively large range of its values that all correspond to the black hole spacetimes. The values of the observers inclination angle $\theta_0$ are chosen to represent the cases of small , mediate, extreme and very extreme inclination in order to fully exhibit the strong effect of $\theta_0$ on the line profile. The radius of radiating ring is restricted to the inner part of Keplerian discs. Again, the three values of $r_e$ reflect the optical effects in the innermost part of the disc, the mediate inner part and those where the effects of the tidal charge become to be effectively suppressed. First, the radii are fixed to be the same in all the spacetimes under consideration ($b=0$, $-1$, $-3$), we have chosen $r_e=6.3M$, $8M$ and $12M$ where the first value ($r_e=6.3M$) correspond to stable circular of spacetime with braneworld parameter value being $b=-3$. The results are presented in Fig.\ref{figure1}. We can see that the influence of the tidal charge is suppressed with growing radius, $r_e$, as can be intuitively expected because the contributions of $b$ fall with $r$ growing faster than those of the mass and spin parameters of the spacetime and this effect is for each inclination angle $\theta_0$. With $r$ approaching the innermost part of the disc, the profiled lines become to be wider and flatter. These effects are strengthened for inclination angle growing, as demonstrated in Fig.\ref{figure2}. Generally,  the profiled lines has two peaks at the red-end and blue-end with the blue-end peak being always higher in comparison with the red one. For fixed $b<0$, the extension, $\Delta g$,  of the profiled line grows with $\theta_0$ growing and $r_e$ falling, while the difference of the profile height at the blue and red end grows with inclination growing. The tidal charge strengthens these effects while going to higher negative values; for example the frequency shift difference $\Delta g$ at the radius $r_e=6.3M$ and the inclination angle $\theta_0=80^\circ$ grows with descending $b$ such that $\Delta g(b=0)=1.3-0.5=0.8$, $\Delta g(b=-1)=1.38-0.5=0.88$ and $\Delta g(b=-3)=1.4-0.4=1.0$.

When the inclination angle reaches extreme values of $\theta_0>85^\circ$, the profiled lines in the innermost part of the stable Keplerian orbits are enriched by humps reflecting strong lensing effects caused by traversing the region in close vicinity of the black hole horizon. The two humps appearing in the profile are caused by the fact that we consider both direct and indirect photons as generating the profiled line.

	The innermost radius $r_e=6.3M$ is chosen to correspond to the marginally stable orbit of K-N spacetime with $a=0.9981$ and $b=-3$. However, for fixed $a$, but $b=-1(0)$, such a radius is far above the related marginally stable orbit at these spacetimes, since $r_{ms}$(a=0.9981, b=-1)=4M while $r_{ms}(a=0.9981,b=0)=1.23M$. Therefore, in order to compare the radiating rings orbiting in close vicinity of the marginally stable orbit assumed to be the edge of a Keplerian disc, we give the profiled lines at $r_e=r_{ms}$ in Fig.\ref{figure3}, for the same inclination angles. Now the effects of the tidal charge are represented in more precise form. Again, the extension of the line $\Delta g$ grows more strongly with descending  $b$. Of course, we can see that for $b=0$, the profiled lines are strongly redshifted and suppressed in magnitude, since for $b=0$ we are dealing with a near extreme spacetime, when the radiating ring is orbiting in extremely deep gravitational potential in close vicinity of the black hole horizon. On the other hand, for $b=-1$ ($b=-3$) the radiating ring is in much higher distance of the horizon and the gravitational field there is not so strong since these black holes (with $a=0.9981$) are far from the extreme black holes.

The lines profiled by relativistic effects in the innermost part of the accretion disc clearly demonstrate the influence of the tidal charge as shown in Fig.\ref{figure3}. The lines become flatter, but again the influence of the inclination angle is quite strong, but brings some characteristic features independently on the value of $b$.

\item $b>=0$.
Now we discuss the case that could correspond also to modelling of profiled lines in the standard Kerr-Newman spacetimes assuming no electromagnetic interaction of the matter of the radiating ring and the black hole electric charge $Q$.

While focusing our attention to the black hole spacetimes, we have to shift the spin parameter $a$ with changes of the tidal charge $b$ (electric charge $Q^2$). The results are shown in Fig.\ref{figure4} for situations studied for the case of $b<0$, i.e., in the close vicinity of $r_{ms}$, in the middle of the innermost part of the disc and in the region where influence of $b$ starts to be highly suppressed. We can see that for large inclination angles  ($\theta_0 \ge 80^\circ$) two humps appear in the line profile, one near the red end of the line, the other close to its blue end.

\end{itemize}

\section{Conclusion}

We have shown that the tidal charge of the braneworld rotating black holes has a general tendency to make profiled lines of the radiating rings in the inner part of an accretion disc to be wider and flatter. We constructed some theoretically grounded quantities based on characteristics of the profiled lines that could be in principle compared with data determining profiled lines observed in microquasars and active galactic nuclei. We have shown that the influence of the inclination angle on the line profiles could be stronger than that of even high values of $b$ (Cf. Fig.\ref{figure1} and \ref{figure2}) that could make the situation difficult for estimating the black hole parameters when we have no limits on the inclination angle, but makes the estimates relatively convincing when we can strongly limit the inclination angle, as we can, at least, for some microquasars. The point is that the inclination angle influence gives some specific features on the profiled lines (e.g. humps for very high $\theta_0>85^\circ$) that could serve as an additional test of the black hole parameters. It is quite important that one such a hump appears in the case of negative braneworld parameters being located near the blue end of the line, while two humps appear for positive braneworld parameters (or equivalently for the Kerr-Newman spacetimes), one near the red end of the line, the other near the blue end of the spectral line. Clearly, all the phenomena deserve attention and bring interesting, principally new results that could be extended for development of methods enabling estimates of the black hole parameters.

\section{Acknowledgements}
This work was supported by the Czech Grant \emph{MSM 4781305903}.



\begin{thebibliography}{10}

\bibitem{Ali-Gum:2005:}
A.~N. Aliev and A.~E. G{\"u}mr{\"u}k{\c c}{\"u}o{\u g}lu.
\newblock Charged rotating black holes on a 3-brane.
\newblock {\em Phys. Review}, 71(10):104027--+, 2005.

\bibitem{Ark-Dim-Dva:1998:}
N.~Arkani-Hamed, S.~Dimopoulos, and G.~Dvali.
\newblock The hierarchy problem and new dimensions at a millimeter.
\newblock 429:263--272, 1998.

\bibitem{Asch:2004:ASTRA:}
B.~Aschenbach.
\newblock Measuring mass and angular momentum of black holes with
  high-frequency quasi-periodic oscillations.
\newblock {\em Astronom. and Astrophys.}, 425:1075--1082, 2004.

\bibitem{Asch:2007:}
B.~Aschenbach.
\newblock Measurement of mass and spin of black holes with qpos.
\newblock {\em ARXIV}, 2007.

\bibitem{Bao-Stu:1992:}
G.~Bao and Z.~Stuchl{\'{i}}k.
\newblock Accretion disk self-eclipse:x-ray light curve and emmision line.
\newblock {\em Astrophys. J.}, 400:163--169, 1992.

\bibitem{Bardeen:1973:}
J.~M. {Bardeen}.
\newblock {Timelike and null geodesics in the Kerr metric.}
\newblock In {\em Black Holes (Les Astres Occlus)}, pages 215--239, 1973.

\bibitem{BS:1976:}
J.~{Bicak} and Z.~{Stuchlik}.
\newblock {On the latitudinal and radial motion in the field of a rotating
  black hole}.
\newblock {\em Bulletin of the Astronomical Institutes of Czechoslovakia},
  27:129--133, 1976.

\bibitem{Carter:1968:}
B.~Carter.
\newblock Global structure of the kerr family of gravitational fields.
\newblock {\em Phys. Rev.}, 174:1559, 1968.

\bibitem{CHANDRA}
S.~Chandrasekhar.
\newblock {\em The Mathematical Theory of Black Holes}.
\newblock Clarendon Press Oxford, Oxford University Press New York, 1983.

\bibitem{Cun-Bar:1973:}
C.~T. Cunningham and J.~M. Bardeen.
\newblock The optical appearance of a star orbiting an extreme kerr black hole.
\newblock {\em The Astrophysical Journal}, 183:237--264, 1973.

\bibitem{Czerny_etal:2007:}
B.~{Czerny}, M.~{Moscibrodzka}, D.~{Proga}, T.~{Das}, and A.~{Siemiginowska}.
\newblock {Low angular momentum accretion flow model of Sgr A* activity}.
\newblock {\em ArXiv e-prints}, 710, October 2007.

\bibitem{Dad-etal:2000:}
N.~Dadhich, R.~Maartens, P.~Papadopoulos, and V.~Rezania.
\newblock Black holes on the brane.
\newblock 487:1, 2000.

\bibitem{Dot:2006}
T.~Dotani.
\newblock Asca and rxte observations of the accretion disk in the x-ray
  binaries.
\newblock In {\em Current High-energy Emission Around Black Holes: Proceedings
  of the 2nd KIAS Astrophysics Workshop : Korea Institute for Advanced Study,
  September 3-8, 2001}, 2002.

\bibitem{Fan-Cal-Fel-Cad:1997:}
C.~Fanton, M.~Calvani, F.~de~Felice, and A.~\v{C}ade\v{z}.
\newblock Detecting accretion disks in active galactic nuclei.
\newblock {\em Publ. Astron. Soc. Japan}, (49):159--169, 1997.

\bibitem{Ger-Maa:2001:}
C.~Germani and R.~Maartens.
\newblock Stars in the braneworld.
\newblock {\em Phys. Review}, 64:124010, 2001.

\bibitem{Ghe-etal:2005:}
A.~M. Ghez, S.~Salim, S.~D. Hornstein, A.~Tanner, J.~R. Lu, M.~Morris, E.~E.
  Becklin, and G.~Duch{\^e}ne.
\newblock Stellar orbits around the galactic center black hole.
\newblock {\em Astrophys. J.}, 620:744--757, 2005.

\bibitem{Yoshi:2007}
Y.~C. Joshi.
\newblock Displacement of the sun from the galactic plane.
\newblock 2007.

\bibitem{Laor:1991:}
A.~Laor.
\newblock Line profiles from a disk around a rotating black hole.
\newblock {\em Astrophys. J.}, 376:90--94, 1991.

\bibitem{Maa:2004:}
R.~Maartens.
\newblock Brane-world gravity.
\newblock {\em Living Rev. Rel.}, 7:7, 2004.

\bibitem{Mat-Fab-Ros:1993:}
G.~Matt, A.~C. Fabian, and R.~R. Ross.
\newblock Iron k-alpha lines from x-ray photoionized accretion discs.
\newblock {\em MNRAS}, 262(1):179--186, 1993.

\bibitem{McCli-Nar-Sha:2007:}
J.~E. McClintock, R.~Narayan, and R.~Shafee.
\newblock Estimating the spins of stellar-mass black holes.
\newblock {\em ArXiv e-prints}, 707, 2007.
\newblock To appear in Black Holes, eds. M. Livio and A. Koekemoer (Cambridge
  University Press), in press (2008).

\bibitem{MTW}
C.~W. Misner, K.~S. Thorne, and J.~A. Wheeler.
\newblock {\em Gravitation}.
\newblock W.~H. FREEMAN AND COMPANY,~San Francisco, 1973.

\bibitem{Ran-Sun:1999:}
L.~Randall and R.~Sundrum.
\newblock An alternative to compactification.
\newblock 83(23):4690--4693, 1999.

\bibitem{Rau-Bla:1994:}
K.~P. Rauch and R.~D. Blandford.
\newblock Optical caustics in a kerr spacetime and the origin of rapid x-ray
  variability in active galactic nuclei.
\newblock {\em The Astrophysical Journal}, (421):46--68, 1994.

\bibitem{Rem:2005:}
R.~A. Remillard.
\newblock X-ray spectral states and high-frequency qpos in black hole binaries.
\newblock {\em Astro. Nach.}, 326(9), 2005.

\bibitem{Rem-McCli:2006:ARASTRA:}
R.~A. Remillard and J.~E. McClintock.
\newblock X-ray properties of black-hole binaries.
\newblock {\em Ann. Rev. of Astron. and Astrophys.}, 44(1):49--92, September
  2006.

\bibitem{Sasaki:2000:}
M.~{Sasaki}, T.~{Shiromizu}, and K.-I. {Maeda}.
\newblock {Gravity, stability, and energy conservation on the Randall-Sundrum
  brane world}.
\newblock {\em Phys. Rev. D}, 62(2):024008--+, July 2000.

\bibitem{SS:a:RAGTime:2007:Proceedings}
J.~Schee and Z.~Stuchl\'{i}k.
\newblock Optical effects in brany kerr spacetimes.
\newblock In {\em Proceedings of RAGtime 8/9: Workshops on black holes and
  netron stars, Opava, 15--19/19--21 September 2006/2007}, 2007.

\bibitem{SS:b:RAGTime:2007:Proceedings}
J.~Schee and Z.~Stuchl\'{i}k.
\newblock Spectral line profile in brany kerr spacetime.
\newblock In {\em Proceedings of RAGtime 8/9: Workshops on black holes and
  netron stars, Opava, 15--19/19--21 September 2006/2007}, 2007.

\bibitem{SSJ:RAGTime:2005:Proceedings}
J.~Schee, Z.~Stuchl\'{i}k, and J.~Jur\'{a}\v{n}.
\newblock Light escape cones and raytracing in kerr geometry.
\newblock In {\em Proceedings of RAGtime 6/7: Workshops on black holes and
  netron stars, Opava, 16--18/18--20 September 2004/2005}, 2005.

\bibitem{Shi-Mae-Sas:2000:}
T.~Shiromizu, K.~Maeda, and M.~Sasaki.
\newblock The einstein equations on the 3-brane world.
\newblock {\em Phys. Review}, 62:024012, 2000.

\bibitem{Strohmayer:2007:}
T.~{Strohmayer}.
\newblock {Understanding the nature of high inclination low mass X-ray
  binaries: broad-band and line spectra from A1744-361}.
\newblock In {\em Chandra Proposal}, pages 2377--+, September 2007.

\bibitem{Stu:1981b:}
Z.~Stuchl{\'{i}}k.
\newblock Null geodesics in the kerr-newman metric.
\newblock {\em Bull. Astron. Inst. Czechosl.}, 32(6), 1981.

\bibitem{Stu-Bao:1992:}
Z.~Stuchl{\'{i}}k and G.~Bao.
\newblock Radiation from hot spots orbiting an extreme reissner-nordström
  black hole.
\newblock {\em General Rel. and Grav.}, 24(9), 1992.

\bibitem{Stu-Kot:2007:}
Z.~Stuchl\'{i}k and A.~Kotrlov\'{a}.
\newblock Orbital resonance model of qpos\\in braneworld kerr black hole
  spacetimes.
\newblock In {\em Proceedings of RAGtime 8/9: Workshops on black holes and
  netron stars, Opava, 15--19/19--21 September 2006/2007}, 2007.

\bibitem{Tor:2005a:}
G.~T\"{o}r\"{o}k.
\newblock A possible 3:2 orbital epicyclic resonance in qpo frequencies of sgr
  a*.
\newblock {\em Astron. and Astrophys.}, 1(440), 2005a.

\bibitem{Tor:2005b:}
G.~T\"{o}r\"{o}k.
\newblock Qpos in microquasars and sgr a* measuring the black hole spin.
\newblock {\em Astronom. Nachr.}, 856(326), 2005b.

\bibitem{Tor-Abr-Klu-Stu:2005:}
G.~T\"{o}r\"{o}k, M.~Abramowicz, W.~Klu\'{z}niak, and Z.~Stuchl\'{i}k.
\newblock {\em Astron. and Astrophys.}, 1(436), 2005.

\bibitem{Viergutz:1993:}
S.~U. {Viergutz}.
\newblock {Image generation in Kerr geometry. I. Analytical investigations on
  the stationary emitter-observer problem}.
\newblock {\em Astronomy and Astrophysics}, 272:355--+, May 1993.

\bibitem{Zak:2003:}
A.~F. Zakharov.
\newblock "the iron $k_{\alpha}$-line as a tool for analysis of black hole
  characteristics".
\newblock {\em Publications of the Astronomical Observatory of Belgrade},
  76:147--162, 2003.

\bibitem{Zak-Rep:2006:}
A.~F. Zakharov and S.~V. Repin.
\newblock "different types of fe $k_{\alpha}$ lines from radiating annuli near
  black holes".
\newblock {\em New Astr.}, 11:405--410, 2006.

\end{thebibliography}

\end{document}